\documentclass[preprint]{elsarticle}

\usepackage{units}
\usepackage{graphicx}
\usepackage[tight]{subfigure}
\usepackage{amsmath,amssymb}
\usepackage{booktabs}

\journal{Mathematics and Computers in Simulation}

\begin{document}

\begin{frontmatter}

\title{Nonparametric Edge Detection in Speckled Imagery}

\author[de]{Edwin Gir\'on}
\ead{eggirona@gmail.com}
\author[ufal]{Alejandro C.\ Frery\corref{cor1}}
\ead{acfrery@gmail}
\ead[url]{http://sites.google.com/site/acfrery}
\author[de]{Francisco Cribari-Neto}
\ead{cribari@de.ufpe.br}
\ead[url]{http://www.de.ufpe.br/~cribari}
\cortext[cor1]{Corresponding author}
\address[de]{Departamento de Estat\'{\i}stica, Universidade Federal de Pernambuco, Cidade Universit\'{a}ria, Recife/PE, 50740--540, Brazil}
\address[ufal]{CPMAT \& LCCV, Instituto de Computa\c c\~ao, Universidade Federal de Alagoas,\\
BR 104 Norte km 97, Macei\'o/AL, 57072--970, Brazil}

\begin{abstract}
We address the issue of edge detection in Synthetic Aperture Radar imagery. In particular, we propose nonparametric methods for edge detection, and numerically compare them to an alternative method that has been recently proposed in the literature. Our results show that some of the proposed methods display superior results and are computationally simpler than the existing method. An application to real (not simulated) data is presented and discussed. 
\end{abstract}


\end{frontmatter}

\section{Introduction}

Synthetic Aperture Radar (SAR) images are an important source information in many applications, such as urban planning, environmental monitoring, crop management, oil prospection, mining exploration, wind detection, animal life detection, among others.
A SAR is a coherent radar of high resolution that works on-board using a synthetic antenna of a movable platform, like an airplane or a satellite, covering extended surfaces and producing images.
SAR systems employ the Doppler effect and processes the signal obtaining high spatial resolution in the direction of the platform motion~\citep{RemoteSensingImagingSar}.

During the data collection, the target remains illuminated under the antenna beam for a few moments and is observed by the radar from positions induced by its movement throughout the platform trajectory.
The radar illuminates the target with a succession of pulses of a given frequency.
The energy is propagated in all directions, and part of it returns to the antenna (this return is called `echo').
The sensor measures both the intensity and the delay between the signals sent from and received back by the antenna. The image is then formed based on the energy returned by each point on the surface.

Some of the most important features of the SAR sensor for remote sensing are:
\begin{itemize}

\item The active nature of the instrument makes it independent of other illumination sources, being able to gather images at any time.

\item Microwaves intensity are not significantly affected by the presence of clouds, so image acquisition is possible in most metheorological conditions and in regions with permanent cloud coverage.

\item SAR images can have high spatial resolution, e.g.\ of less than one meter, thus making the study of small scale phenomena possible.

\item SAR images contain complementary information to that provided by optical images. The selection of frequency band, polarization and angle of incidence in SAR imagery allows the discrimination of different surface properties.

\end{itemize}

A SAR sensor emits and receives electromagnetic waves of complex nature and, therefore, the received signal can be stored in different formats: complex, intensity, amplitude and phase \citep{Oliver1998}.

In a SAR image, it is possible to distinguish several types of roughness or texture, according to which one can classify the different types of covers:
\begin{itemize}

\item Homogenous areas: Surfaces of very little texture; for example, crops, deforestation, and, under some conditions, snow, water or ice.

\item Heterogeneous areas: Surfaces that display some texture; for example, forests on not very pronounced reliefs, among others.

\item Extremely heterogeneous areas: Surfaces with intense texture; for instance, urban areas, among others.

\end{itemize}

``Texture'', in the context of SAR imagery, should be understood as a measure of the number of objects in a cell of the size the of the wavelength employed by the sensor.
A fine texture corresponds to a large number of objects per cell, while coarse or extremely heterogeneous textures are those for which only a few objects are counted per cell.
The Japanese Earth Resources Satellite JERS-1, for instance, operates on L-Band (\unit[$1.3$]{GHz}, \unit[$23.5$]{cm} wavelength) and the European Remote Sensing Satellites ERS-1 and ERS-2 use C-Band (\unit[$5.3$]{GHz}, \unit[$5.6$]{cm}  wavelength).

A problem of paramount importance in the analysis of images is segmentation: the process that divides an image in its constituent parts or objects.
Its main goal is to group image areas that have similar characteristics.
One of the basic principles in the segmentation process is the detection of discontinuities.
\textit{Edges} are the borders of the objects and are therefore quite useful for their segmentation, registration and identification.
Edges can be thought as the locations where abrupt changes in intensity or in other important characteristic occur.

The quality of SAR images is degraded by speckle, a degradation which follows from the use of coherent illumination, i.e., when the signal phase is employed in the image formation.
Such degradation is characteristic of technologies that employ microwaves, sonar, laser and ultrasound.

The presence of speckle makes edge detection difficult, since most algorithms identify regions using local characteristics.
Though speckle should not be regarded as noise, since it has a deterministic nature and is reproducible, from the image practitioner viewpoint it can be considered a random effect and can be conveniently described by stochastic laws; c.f.\  \citet[Sec.~4.3.1]{RemoteSensingImagingSar}.
It is not convenient to only use pointwise information when detecting edges under speckle; it is necessary to analyze the image using sets of pixels that provide local information \citep{Gambini2006}.

Different approaches can be used to locate the edges between regions in a SAR image.
A particularly attractive and well performing statistical method was proposed by \citet{Gambini:StatisticsComputing}.
It is based on the family of $\mathcal{G}$ distributions, which can be successfuly used to describe areas with different degrees of homogeneity~\citep{FreitasFreryCorreia:Environmetrics:03,Frery1997,Mejail2001,Mejail2003}.

In this work we consider intensity imagery, described by the $\mathcal{G}^0_I$ law.
This distribution is indexed by the number of looks $L\geq1$, the scale parameter $\gamma>0$ and the roughness parameter $\alpha<0$.
The former can be controlled when generating the image or in postprocessing stages, and is a measure of the signal-to-noise ratio~\cite[Sec.~4.3.1]{RemoteSensingImagingSar}.
The value of the roughness parameter is of interest in many applications, since it can be used as an indicator of land type.
The scale parameter relates to the relative power between the reflected and incident signals~\citep{Frery1997}.

\begin{figure}[hbt]
\centering
\includegraphics[width=.8\linewidth]{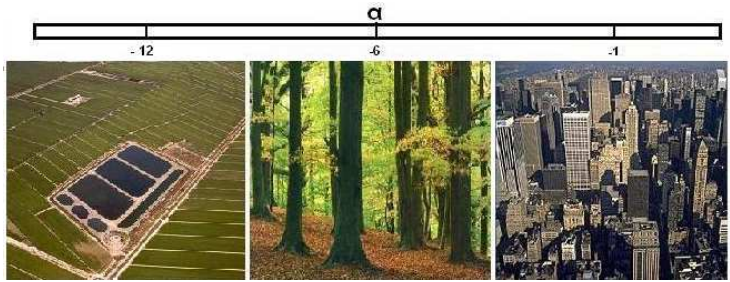}
\caption{Different targets and associated roughness parameter values}\label{interpret_alpha}
\end{figure}

Figure~\ref{interpret_alpha} presents three different targets and the corresponding values (or range of values) of the roughness parameter ($\alpha$).
Small values of $\alpha$  (e.g., $\alpha<-10$) are associated with homogeneous areas, such as pastures.
Values of $\alpha\in[-10, -4]$ are characteristic of heterogeneous regions, for example forests.
Finally, larger values of $\alpha$ (say, $-4<\alpha<0$) are observed in extremely heterogeneous areas, such as urban regions~\citep{Bustos2002,Mejail2003}.

Our chielf goal is to develop and assess new methods for edge detection in SAR images using computationally efficient nonparametric statistical inference.
\citet{Gambini:StatisticsComputing} showed that a method based on maximum likelihood, which is presented in Section~\ref{sec:Gambini}, is more precise than four commonly used techniques: two based on raw data (maximum discontinuity and fractal dimension) and two based on estimates (maximum discontinuity and anisotropic smoothed roughness).
Our numerical evaluation reveals that some of the edge detection strategies we propose in this paper outperform the best method available to date, i.e., maximum likelihood~\citep{Gambini:StatisticsComputing}, at considerably lower computational cost.

The paper unfolds as follows.
Section~\ref{chap:model} presents the model.
Section~\ref{chap:method} discusses the use of nonparametric statistics for SAR image detection, while Section~\ref{chap:results} presents the results. An application that uses real (not simulated) data is presented and discussed in Section~\ref{chap:application}. 
Conclusions and directions for future work are outlined in Section~\ref{chap:conclu}.

\section{The multiplicative model}\label{chap:model}

The physics of SAR image formation leads to the multiplicative model: the random field $Z$ describing the observations can be viewed as the product of two independent random fields, which are not observed directly, namely: $X$ and $Y$.
The former ($X$) models the properties of the imaged area (backscatter), whereas the latter ($Y$) models the speckle noise due to the use of coherent illumination.

Speckle noise in intensity $L$-looks format follows a Gamma distribution, denoted by  $Y \thicksim \Gamma(L,L)$, whose density is given by
\begin{equation*}
f_{Y}(y)= \frac{L^L}{\Gamma(L)}y^{L-1} \exp\{-Ly\}, \qquad L\geq 1, y > 0.
\label{distribution_Y}
\end{equation*}

In SAR images modeling, the smallest value of $L$ is $1$, which corresponds data with the highest spatial resolution, at the expense of lower signal-to-noise ratio.
The number $L$ can be assumed to be known or estimated beforehand from the entire image.
We shall assume that the number of looks is known.

The backscatter exhibits different degrees of homogeneity, and different models have been used to encompass this characteristic.
The reciprocal of Gamma distribution is a useful and tractable model \citep{Allende2006,Bustos2002,Cribari-Neto2002,Frery2004,Frery1997,Gambini:StatisticsComputing,Gambini2006,Mejail2001,Mejail2003,Moschetti2006,NascimentoCintraFrertIEEETGARS,Vasconcellos2005}.
The random variable $X$ follows this law, denoted $X \thicksim \Gamma^{-1} (\alpha, \gamma)$, if its density function is
\begin{equation*}
f_X(x)=\frac{1}{\gamma^{\alpha}\Gamma(-\alpha)}x^{-\alpha-1} \exp\Bigl\{-\frac{\gamma}{x}
\Bigr\}, \qquad -\alpha, \gamma, x>0.
\end{equation*}

Let $X \thicksim \Gamma^{-1}(\alpha, \gamma)$  and $Y \thicksim \Gamma(L, L)$ be two independent random variables. It can be shown that the random variable $Z = X Y$ follows a $\mathcal{G}^0_{\mathcal{I}}$ distribution, denoted $Z \thicksim \mathcal{G}^0_{\mathcal{I}} (\alpha, \gamma, L)$, whose density is 
\begin{equation}\label{fun_g0i}
f_{Z}(z)=\frac{L^L \Gamma(L-\alpha)}{\gamma^{\alpha} \Gamma(L)
\Gamma(-\alpha)}\frac{z^{L-1}}{(\gamma + Lz)^{L-\alpha}}, \qquad z>0,
\end{equation} where $-\alpha> 0$ is the roughness parameter, $\gamma> 0$ is the scale parameter and $L \geq 1$ is the number of looks~\citep{Frery1997}.

In this work we consider independent samples.
Correlated fields are treated by \citet{SimulationSpatiallyCorrelatedClutter2009}.

The $r$-th noncentral moment of $Z$ is given by
\begin{equation}\label{Ez}
\operatorname{E}[Z^r]=\left(\frac{\gamma}{L} \right)^{r} \frac{\Gamma(-\alpha
-r)\Gamma(L+r)}{\Gamma(-\alpha)\Gamma(L)},
\end{equation}
if $-\alpha>r$, and $\infty$ otherwise.

The $\mathcal{G}^0_{\mathcal{I}}$ distribution is very attractive for modeling data with speckle noise, due to its mathematical tractability and because it is able to describe information from most types of areas.

\begin{figure}[hbt]
\centering
\includegraphics[width=.8\linewidth, bb = 60 180 540 560]{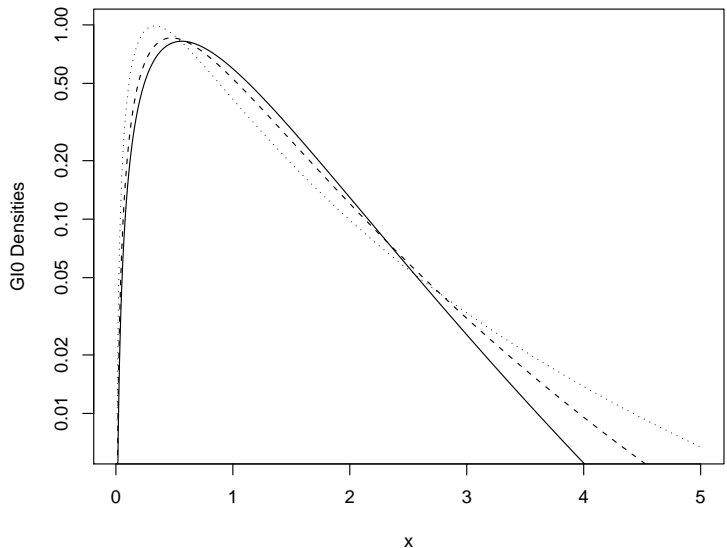}
\caption{$\mathcal{G}^0_{\mathcal{I}}(\alpha,\gamma_{\alpha,3},3)$ densities for $\alpha =-3, -6, -12$ (dots, dashes, solid)}\label{goi_alfas}
\end{figure}

Figure~\ref{goi_alfas} shows $\mathcal{G}^0_{\mathcal{I}}(\alpha, \gamma_{\alpha,3}, 3)$ densities, where $\gamma_{\alpha,L}$ is the value of $\gamma$ that delivers unit expected value for given $\alpha<-1$ and $L$.
The densities are presented in semilogarithmic scale, showing that they have heavy (linear) tails with respect to the Gaussian distribution which displays quadratic behavior.
It is noticeable that larger values of $\alpha$ lead to larger variances; in fact, the variance is not finite when $\alpha\geq -1$.

\begin{figure}
\centering
\subfigure[Single look]{\includegraphics[bb = 0 0 360 360, width=.32\linewidth]{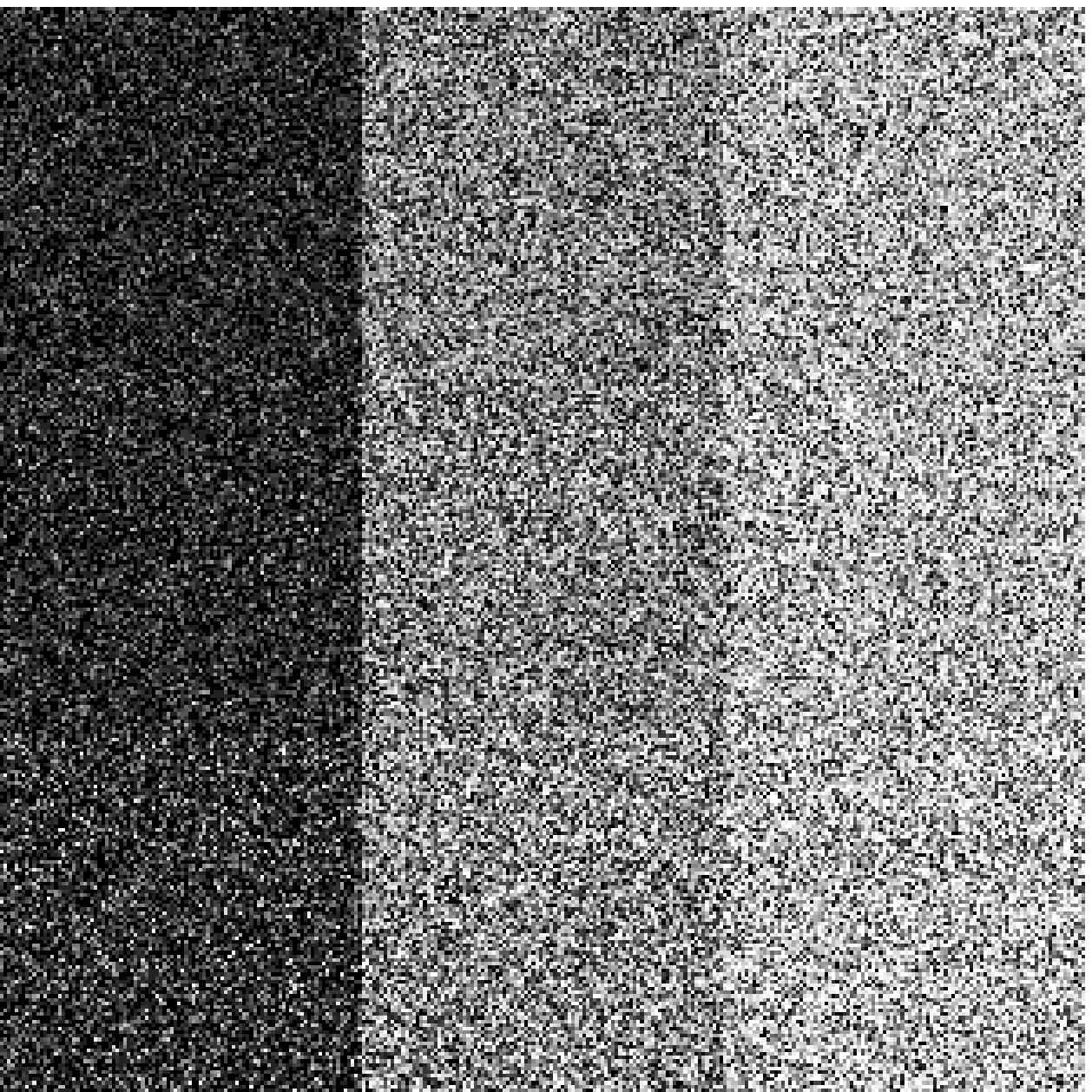}}
\subfigure[Three looks]{\includegraphics[bb = 0 0 360 360, width=.32\linewidth]{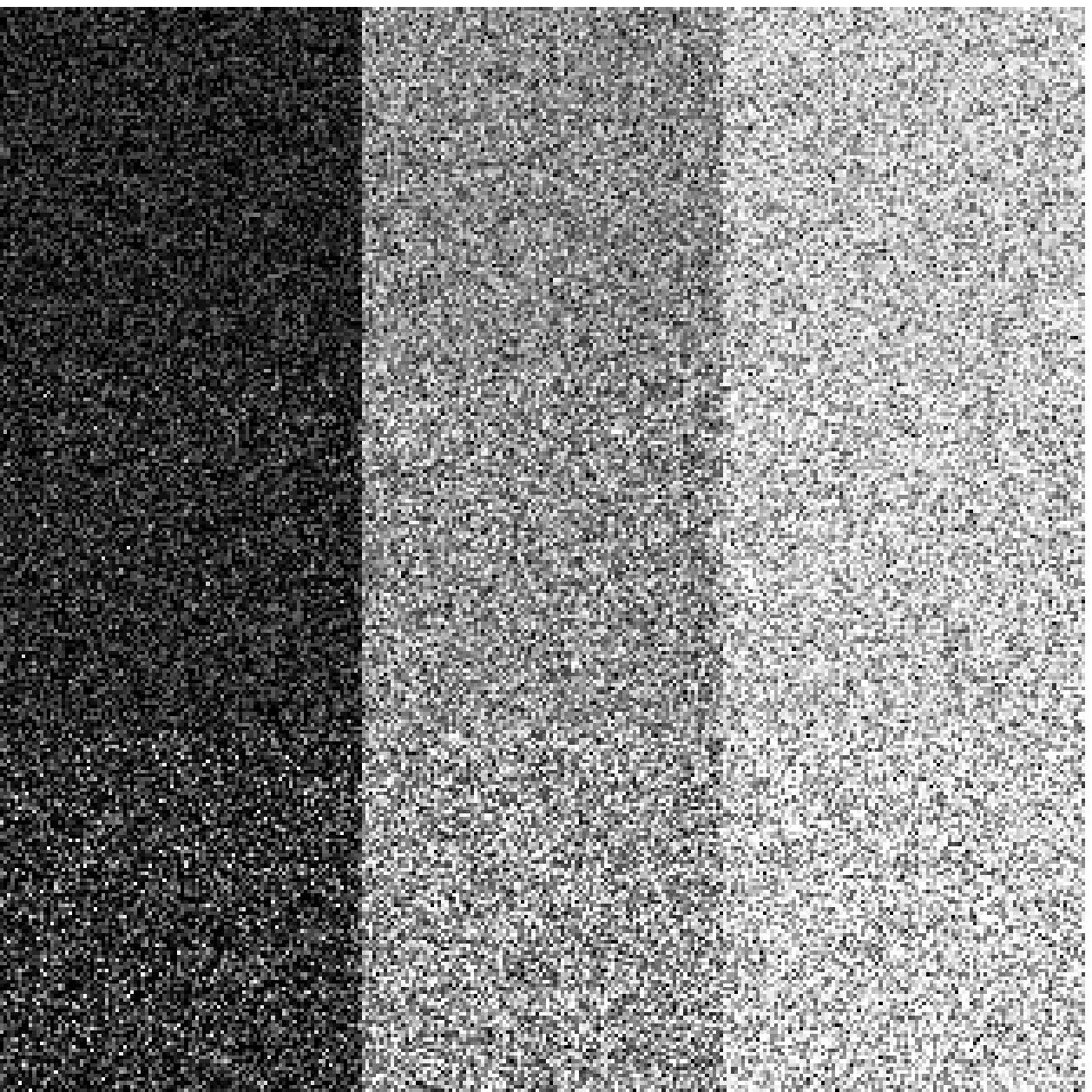}}
\subfigure[Eight looks]{\includegraphics[bb = 0 0 360 360, width=.32\linewidth]{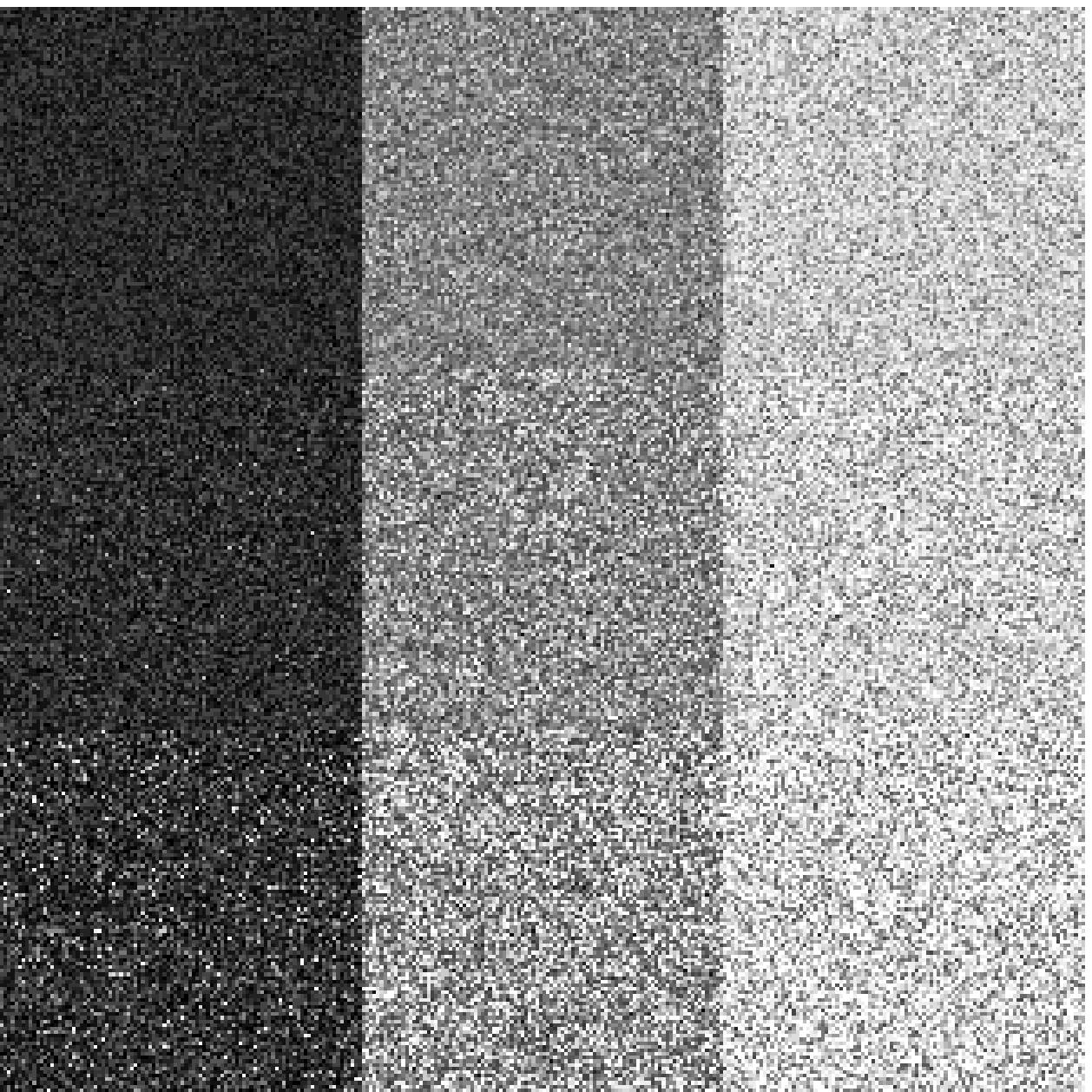}}
\caption{Patches of speckled data}\label{fig:patches}
\end{figure}

Figure~\ref{fig:patches} shows nine patches of speckled data for one, three and eight looks.
Each patch consists of $3\times3$ images with varying roughness and contrast: each column shows images with same mean ($1$, $5$ and $10$, left to right) while rows from bottom to top show images with $\alpha\in\{-2,-5,-20\}$.
It can be readily seen that finding edges in speckled data can be a hard task due to the existing local variation.

\citet{Gambini:StatisticsComputing,Gambini2006} used an analogy estimator based on moments of order $1/2$ and $1$.
Since we wish to extend their work, we shall use the same approach.

Let $(Z_1 ,\dots, Z_n )$ be a vector of independent identically distributed random variables, with common distribution $\mathcal G_{\mathcal I}^0(\alpha,\gamma,L)$, $\alpha<-1/2$, $\gamma>0$ and $L$ known.
Define the $r$-th sample moment as
\begin{equation*}
\widehat{m}_r=\frac{1}{n}\sum_{i=1} ^{n}z^r_i,
\end{equation*}
with $r=1/2$ and $1$ (that is the reason why we assume $\alpha<-1$ rather than $\alpha<0$).

From equation~\eqref{Ez} it is immediate that
$$
\operatorname{E}(Z)=\left(\frac{\gamma}{L} \right) \frac{\Gamma(-\alpha
-1)\Gamma(L+1)}{\Gamma(-\alpha)\Gamma(L)}, \qquad -\alpha>1,
$$
and that
$$\operatorname{E}(Z^{\frac12})= \left(\frac{\gamma}{L} \right)^{\frac12} \frac{\Gamma(-\alpha
-\frac12)\Gamma(L+\frac12)}{\Gamma(-\alpha)\Gamma(L)}, \qquad -\alpha>1/2.
$$

Replacing the population moments by their sample counterparts, and the parameters by the corresponding estimators, we arrive at the following system of two equations:
\begin{equation}\label{parametro_gamma}
m_1=\left(\frac{\widehat{\gamma}}{L}
\right)\frac{\Gamma(-\widehat{\alpha}
-1)\Gamma(L+1)}{\Gamma(-\widehat{\alpha})\Gamma(L)}, \qquad
-\widehat{\alpha}>1, 
\end{equation}
and
\begin{equation*}
m_{\frac{1}{2}}=\left(\frac{\widehat{\gamma}}{L}
\right)^{\frac{1}{2}} \frac{\Gamma(-\widehat{\alpha}
-\frac{1}{2})\Gamma(L+\frac{1}{2})}{\Gamma(-\widehat{\alpha})\Gamma(L)}, 
\qquad -\widehat{\alpha}>1/2,
\end{equation*}
which leads to the following equation that can be solved numerically in order to obtain an estimator for $\alpha$:
\begin{equation*}
\frac{m_1
\Gamma(-\widehat{\alpha})\Gamma(L)L}{\Gamma(-\widehat{\alpha}
-1)\Gamma(L+1)} = \frac{m^2_{\frac{1}{2}}
\Gamma^2(-\widehat{\alpha})\Gamma^2(L)L}{\Gamma^2(-\widehat{\alpha}
-\frac{1}{2})\Gamma^2(L+\frac{1}{2})}.
\end{equation*}
By plugging the value of $\widehat{\alpha}$ into equation~\eqref{parametro_gamma} we obtain $\widehat{\gamma}$.

\citet{Frery2004} showed that computing ML estimators for the $\mathcal G^0$ family is prone to severe numerical instabilities, and they proposed an iterative algorithm that alleviates this problem.
Recently \citet{DealingMonotoneLikelihood} analyzed this issue and found that it is related to a flattening of the likelihood function, and they proposed a correction based on resampling.
\citet{Cribari-Neto2002} showed that ML estimators for that distribution can be quite biased, and they evaluated the effectiveness of improving them by several resampling techniques.
\citet{Vasconcellos2005} proposed an analytical bias correction for ML estimators; they showed that there is a wide range of practical situations for which the corrected estimator effectively reduces both bias and mean square error of the original ML estimator.
\citet{Bustos2002} derived M-estimators and showed that such robust estimators are superior to the classical estimators in the presence of corner reflectors, a common source of contamination in SAR images.
\citet{Allende2006} derived AM-estimators (M estimators with asymmetric influence functions), motivated by the shape of the $\mathcal{G}^0_A$ density. 
Overall, their estimators outperform both ML and M-estimators.

In the next section we shall summarize the main techniques that are available for edge detection in SAR imagery, with special emphasis on those that explicitly employ statistical models and techniques.

\section{Edge detection in SAR imagery}\label{chap:method}

Statistical edge detection is described in \citet{BOVIK1986}.
These authors introduced nonparametric statistics for edge detection under Gaussian additive noise.
They showed the usefulness of the median and the Wilcoxon-Mann-Whitney tests for edge detection with the help of a sample image interpreted visually.

\citet{Fesharaki1994} proposed an algorithm for edge detection by using a $t$-test, while \citet{Beauchemin1998} used a nonparametric alternative based on the Wilcoxon-Mann-Whitney statistics for detecting changes between adjacent pixel neighborhoods.
Although these tests may be appropriate for specific types of images, they may not detect changes in local grey level values in images with low signal-to-noise ratios, as is the case of SAR imagery.

\citet{HoonLim} compared two-sample tests for edge detection in noisy images.
Later \citet{Lim2006} described a new edge detector based on the robust rank-order test, an alternative to the Wilcoxon test, using $\ell \times \ell$ windows for detecting all possible edges in noisy images.
This method is based on testing whether an $\ell \times \ell$ window is partitioned into two sub-regions.

The detection of edges in images with speckle noise has been studied by many authors, and a variety of techniques are presented and compared by \citet{Gambini:StatisticsComputing,Gambini2006}, whose proposed approach outperforms all competing procedures with an acceptable computational cost.
Their method is based on fitting contours of objects and regions using B-splines; such curves depend on a few parameters and can be easily computed from control points and smoothness conditions~\citep{Brigger2000}.

\citet{Gambini:StatisticsComputing,Gambini2006} used amplitude data for finding edges between different regions.
Such data, if squared, follow the $\mathcal{G}^0_{\mathcal{I}}$ distribution.
Under this model, the regions of the image with different degrees of homogeneity are characterized by the parameters of the distribution.
If a point belongs to the edge of the object, then a sudden change in the parameter values is expected in its neighborhood.

In what follows we present five techniques for detecting edges in SAR images, namely the method proposed by \citet{Gambini:StatisticsComputing,Gambini2006}, and four alternative techniques based on ranks.
The two main advantages of nonparametric tests are that
\begin{enumerate}
\item they do not depend upon the data distribution, and
\item they are robust to extreme observations.
\end{enumerate}
These features are of particular interest in the case at hand, namely, speckled imagery.

\subsection{The Gambini algorithm for edge detection}\label{sec:Gambini}

This is an iterative procedure that refines an initial region until the final result is achieved.
It is based on the fact that if a point belongs to the object edge, then a sample taken from its neighbourhood should exhibit a change in the parameter values and, therefore, should be considered a transition point.

\begin{figure}[hbt]
\centering
\includegraphics[height=6cm, width=8cm]{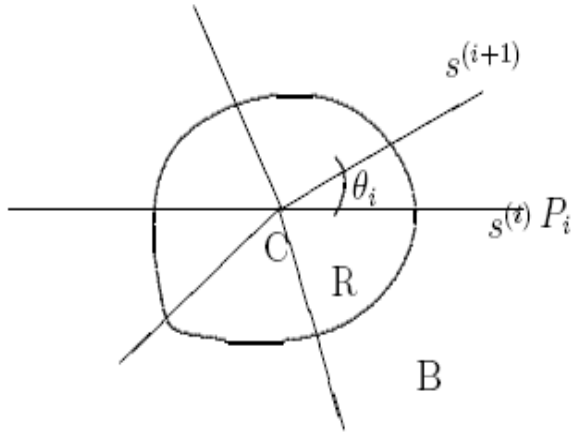}
\caption[Radial straight lines projected]{Radial lines from the
centroid $C$ to the exterior of the region, with a separation of
$\theta_i$}\label{cp_gam}
\end{figure}

Consider $N$ image segments, $s^{(i)}, i\in \{1,\dots, N \}$, of the form $s^{(i)} = \overline{CP_i}$, where $C$ is the centroid of the initial region, the extreme $P_i$ is a point outside the region and $\theta_i= \angle(s^{(i)}, s^{(i+1)})$  is the angle between two successive segments, as shown in Figure~\ref{cp_gam}.

\begin{figure}[hbt]
\centering
\includegraphics[width=0.6\linewidth]{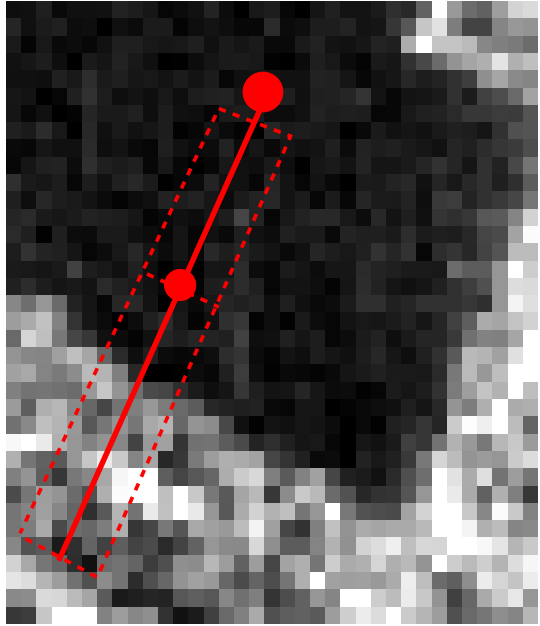}
\caption{Segment, strip and candidate edge point (small red dot)}\label{fig:ShowEdge}
\end{figure}

Consider a strip of pixels around each segment $s^{(i)}$, as illustrated in Figure~\ref{fig:ShowEdge}, partitioned into two areas by a candidate edge point $c_i$ (the small red dot).
The observations within the strip are taken to come from two different models, namely one corresponding to the object $\mathcal{G}^0_{\mathcal{I}}(\alpha_\ell, \gamma_\ell, L)$ (the dark area in Figure~\ref{fig:ShowEdge}) and the other from the background $\mathcal{G}^0_{\mathcal{I}}(\alpha_r, \gamma_r, L)$.

The parameters $(\alpha_\ell, \gamma_\ell)$ and $(\alpha_r, \gamma_r)$ index the region and its background, respectively, and their estimation is described in Section~\ref{chap:model}; note that the estimators depend upon the the transition point $c_i$.

In order to find the transition point on each segment $s^{(i)}$, an objective function is considered: the sample likelihood, which is given by
\begin{equation*}
\ell({\alpha_\ell}, {\gamma_\ell},{\alpha_r}, {\gamma_r})
=\prod^{j}_{i=1}\Pr(z_i; {\alpha_\ell},
{\gamma_\ell})\times \prod^{m}_{i=j+1}\Pr(z_i;
{\alpha_r}, {\gamma_r}),
\end{equation*}
where $j$ is the number of observations in the strip around segment $s^{(i)}$ lying between $C$ and $c_i$, and $m$ is the number of observations in the strip lying between $c_i$ and $P_i$.

\begin{figure}[hbt]
\centering
\includegraphics[height=6cm, width=10cm]{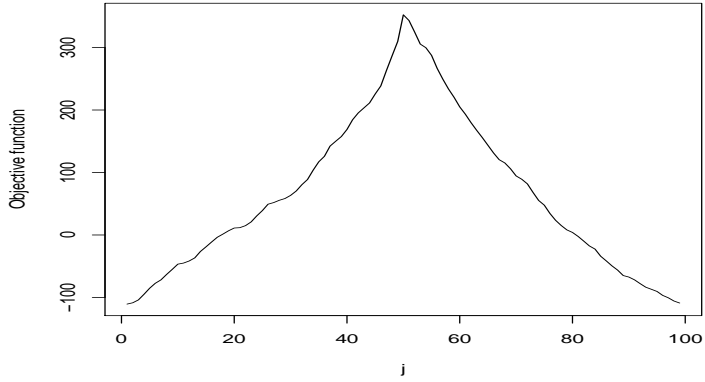}
\caption{Values of the objective function for a
segment of straight}\label{func_vero}
\end{figure}

In order to find the transition point, we maximize the log-likelihood function
\begin{equation*}
\mathcal L=\ln(\ell)=\sum^{j}_{i=1}\ln f_{\mathcal{G}^0_{\mathcal{I}}}(z_i; {\alpha_\ell},
{\gamma_\ell}) + \sum^{m}_{i=j+1}\ln f_{\mathcal{G}^0_{\mathcal{I}}}(z_i; {\alpha_r},
{\gamma_r})
\end{equation*}
for every possible value of $j$.
Using equation~\eqref{fun_g0i}, and assuming that $L$ is known,
\begin{align}
\mathcal L=&\sum^{j}_{i=1}\ln \frac{L^L
\Gamma(L-\widehat{\alpha_\ell})z_i^{L-1}}{\widehat{\gamma_{\ell}}^{\widehat{\alpha_\ell}}\Gamma(L)\Gamma(-\widehat{\alpha_\ell})(\widehat{\gamma_\ell}
+ L z_i)^{L-\widehat{\alpha_\ell}}} + \nonumber\\
& + \sum^{m}_{i=j+1} \ln \frac{L^L
\Gamma(L-\widehat{\alpha_r})z_i^{L-1}}{\widehat{\gamma_{r}}^{\widehat{\alpha_r}}\Gamma(L)\Gamma(-\widehat{\alpha_r})(\widehat{\gamma_r} + Lz_i)^{L-\widehat{\alpha_r}}} .\label{eq:logver}
\end{align}
Finally, the estimated transition point on the segment is given by
\begin{equation}\label{max_j}
\widehat{\jmath}= \arg\max_j \mathcal L.
\end{equation}
Figure~\ref{func_vero} shows typical values of the objective function, taken along a straight line segment.

\subsection{Nonparametric edge detection}

In the typical two-sample situation, the experimenter has two samples from possibly different populations, and wishes to use a statistical test to determine whether the null hypothesis that the two populations are identical should be rejected.
That is, the experimenter wishes to identify differences between the two populations on the basis of their random samples.

An intuitive approach to the two-sample problem is to combine both samples into a single ordered sample, and assign ranks to the observations, regardless the source population.
A possible test statistic is the sum of the ranks assigned to each population.
If such a sum is small (large), there is indication that the values from that population tend to be smaller (larger) than the values obtained from the other population.
The null hypothesis should be rejected if the sum of the ranks associated with one sample is considerably larger than the other sum.

\subsubsection{The Mann-Whitney test}

The data consist of samples from two populations.
Let $X_1,X_2,\ldots,X_n$ denote the random variables sampled from population $A$ and $Y_1,Y_2,\ldots,Y_m$ denote the variates obtained from population $B$.
Assign the ranks $1,\dots,N=n+m$ to the individual observations in the combined sample.
Let $R(X_i)$ and $R(Y_j)$ denote the ranks assigned to $X_i$ and $Y_j$ for all $i$ and $j$.
If two or more sample values are equal (ties), assign to each the average of the ranks that would have been assigned had there been no ties.

The following assumptions are made:
\begin{itemize}
\item Both samples are random samples from the respective populations.
\item In addition to independence within each sample, there is mutual independence between the two samples.
\item The measurement scale is at least ordinal.
\end{itemize}

Let $F(x)$ and $G(x)$ be the distribution functions of $X$ and $Y$, respectively. We wish to test 
\begin{align*}
\mathcal H_0&:F(x)=G(x) \\
\mathcal H_1&:F(x)\neq G(x).
\end{align*}

In many situations, as in edge detection, differences between distributions imply that $\Pr(X<Y)$ does not equal $1/2$. Therefore, we can rewrite the above hypotheses as 
\begin{align*}
\mathcal H_0&:\Pr(X<Y)=\frac{1}{2} \\
\mathcal H_1&:\Pr(X<Y)\neq \frac{1}{2}.
\end{align*}

The null hypothesis can be tested using the Mann-Whitney test, which is unbiased and consistent. The test statistic is computed as follows. When there are no or just a few ties, the sum of the ranks assigned to the sample from population $A$ can be used as a test statistic: 
\begin{equation*}
T=\sum_{i=1}^n R(X_i).
\end{equation*}
When the number of ties is large, one can subtract the (null) mean from $T$ and divide the resulting difference by the (null) standard deviation to get
\begin{equation*}
T_1=\frac{T-n \frac{N+1}{2}}{\sqrt{\frac{nm}{N(N-1)}\sum_{i=1}^N
R_i^2 - \frac{nm(N+1)^2}{4(N-1)} }}.
\end{equation*}

\subsubsection{The Kruskal-Wallis test}

The Mann-Whitney test for two independent samples was extended to the problem of analyzing $k$ independent samples, $k\geq2$, by Kruskal and Wallis in 1952.
The experimental situation is that $k$ random samples have been obtained from $k$ possibly different populations, and one wishes to test the null hypothesis that all populations are identical. 
The Kruskal-Wallis test statistic is a function of the ranks of the observations in the combined sample (like the Mann-Whitney test statistic).

The data consist of $k$ random samples of possibly different sizes.
Denote the $i$th random sample of size $n_i$ by $X_{i1},X_{i2},\ldots,X_{in_i}$. Then, the data may be arranged in columns:
\begin{center}
\begin{tabular}{cccc}
Sample $1$ & Sample $2$ & $\cdots$ & Sample $k$ \\
$X_{1,1}$ & $X_{2,1}$ & & $X_{k,1}$ \\
$X_{1,2}$ & $X_{2,2}$ & & $X_{k,2}$ \\
$\vdots$ & $\vdots$ & $\ddots$ & $\vdots$ \\
$X_{1,n_1}$ & $X_{2,n_2}$ & & $X_{k,n_k}$ \\
\end{tabular}
\end{center}

Let $N$ denote the total number of observations, i.e., $N=\sum_{i=1}^k n_i$.
Assign rank $1$ to the smallest of the $N$ observations, rank $2$ to the second smallest observation, and so on. 
Let $R_i$ be the sum of the ranks assigned to the $i$th sample:
\begin{equation}\label{Ri_KRUSKAL}
R_i = \sum_{j=1}^{n_i} R(X_{ij}) \quad i=1,2, \ldots, k.
\end{equation}
Compute $R_i$ for each sample. Under ties, assign the average rank to each tied observation, as in the previous test.

The following assumptions are made: 
\begin{itemize}
\item All samples are random samples from their respective populations.

\item In addition to independence within each sample, there is mutual independence amongst samples.

\item The measurement scale is at least ordinal.

\item Either the $k$ population distribution functions are identical or else some of the populations tend to yield larger values than the remaining populations.
\end{itemize}

The null and alternative hypotheses are 
\begin{align*}
\mathcal H_0&: \text{All $k$ population distribution functions are identical},\\
\mathcal H_1&: \text{At least one population tends to yield observations larger than}\nonumber\\
&\text{at least one of the remaining populations.}
\end{align*}

Since the Kruskal-Wallis test is designed to be sensitive against differences among means in the $k$ populations, the alternative hypothesis is sometimes stated as 

$\mathcal H_1:$ The $k$ populations do not all have identical means.

The test statistic $T$ is defined as
\begin{equation}\label{t}
T=\frac{1}{S^2}
\left(\sum^k_{i=1}\frac{R^2_i}{n_i}-\frac{N(N+1)^2}{4} \right),
\end{equation}
where 
\begin{equation*}
S^2=\frac{1}{N-1} \left(\sum_{\text{all ranks}}R(x_{ij})^2-\frac{N(N+1)^2}{4} \right).
\end{equation*}

Under no ties, $S^2$ simplifies to $N(N+1)/12$, and the test statistic reduces to
\begin{equation}\label{t_semp}
T_k=\frac{12}{N(N+1)}\sum^k_{i=1}\frac{R^2_i}{n_i} -3(N+1).
\end{equation}
When the number of ties is small or moderate, there is little difference between equations~\eqref{t} and~\eqref{t_semp}, and equation~\eqref{t_semp} is preferred.

\subsubsection{The squared ranks test for variances}

The squared ranks test can be used to assess equality of variances across two or more independent, random samples which have been measured using a scale that is at least interval \citep{Conover}.

The data consist of the two random samples.
Let $X_1,X_2, \ldots, X_n$ denote a random sample of size $n$ from population $A$ and $Y_1,Y_2, \ldots, Y_m$ represent a random sample of the size $m$ from population $B$.
Define 
\begin{equation*}
U_i=|X_i-\widehat{\mu}_1|, \quad i=1,\ldots, n, 
\end{equation*}
and
\begin{equation*}
V_j=|Y_j-\widehat{\mu}_2|, \quad j=1,\ldots, m, 
\end{equation*}
where $\widehat{\mu}_1$ and $\widehat{\mu}_2$ are the two sample means.

Assign ranks $1$ to $n+m$ to the combined sample, as usual.
If several values of $U$ and/or $V$ are equal (ties), assign to each the average of the ranks that would have been assigned to them had there been no ties.

The required assumptions can be stated as follows:
\begin{itemize}
\item Both samples are random samples from the respective populations.

\item In addition to independence within each samples, there is mutual independence between samples.

\item The measurement scale is at least interval.
\end{itemize}

The null and alternative hypotheses are 
\begin{align*}
\mathcal H_0: & \,\, X\text{ and } Y \text{are identically distributed, except for possibly
different means}, \\
\mathcal H_1: & \,\, \operatorname{Var}(X)\neq \operatorname{Var}(Y).
\end{align*}

If there are no values of $U$ tied with values of $V$, the sum of the squared ranks assigned to population $A$ can be used as a test statistic:
\begin{equation*}
T=\sum^n_{i=1}[R(U_i)]^2.
\end{equation*}
If there are ties, subtract the (null) mean from $T$ and divide the difference by the (null) standard deviation to get
\begin{equation*}
T_v=\frac{T-n\overline{R^2}}{\sqrt{\frac{nm}{N(N-1)}\sum^N_{i=1}R_i^4
- \frac{nm}{N-1}(\overline{R^2})^2 }},
\end{equation*}
where $N=n+m$, and $\overline{R^2}$ is the average of the squared ranks of both samples combined:
\begin{equation*}
\overline{R^2}=\frac{1}{N}\left\{ \sum^n_{i=1}[R(U_i)]^2 +
\sum^m_{j=1}[R(V_j)]^2 \right\}; 
\end{equation*}
also, 
\begin{equation*}
\sum^N_{i=1}R_i^4= \sum^n_{i=1}[R(U_i)]^4 + \sum^m_{j=1}[R(V_j)]^4.
\end{equation*}

\subsubsection{The TPE empirical statistic}

The empirical statistic \textbf{TPE} is also based on ranges, and is well suited for situations where one wishes to test whether two samples come from the same distribution or from distributions with different means and/or variances.
The data consist of the two random samples.
Let $X_1,X_2,\ldots,X_n$ denote a random sample of size $n$ from population $A$ and let $Y_1,Y_2,\ldots,Y_m$ denote a random sample of the size $m$ from population $B$.
Assign ranks $1,\dots,N=n+m$, and compute $\overline{X} = n^{-1} \sum_{i=1} R(X)_i$, $\overline{Y} = m^{-1} \sum_{i=1}^ R(Y)_i$, and $D_E = |\overline{X}-\overline{Y}|$.
The mean rank of the combined sample is $\mu = (N +1) / 2$.
The empirical statistic \textbf{TPE} is $E=|D_E - \mu|$.

\subsection{Proposal}

Our chief goal is to perform edge detection in SAR images using the aforementioned nonparametric tests statistics instead of the likelihood function presented in equation~\eqref{eq:logver}.

The new noparametric edge estimates $\widehat{\jmath}$ on segment $s^{(i)}$ are given by
\begin{itemize}
\item Mann-Whitney estimate: $\widehat{\jmath}= \arg\max_j T_{1,j}$,

\item Kruskal-Wallis estimate: $\widehat{\jmath}= \arg\max_j T_{k,j}$,

\item Squared Ranks estimate: $\widehat{\jmath}= \arg\max_j T_{v,j}$, and

\item TPE empirical estimate: $\widehat{\jmath}= \arg\min_j E_j$,
\end{itemize}
where $j$ denotes the edge candidate coordinate.

\begin{figure}[hbt]
\centering
\includegraphics[width=0.45\linewidth]{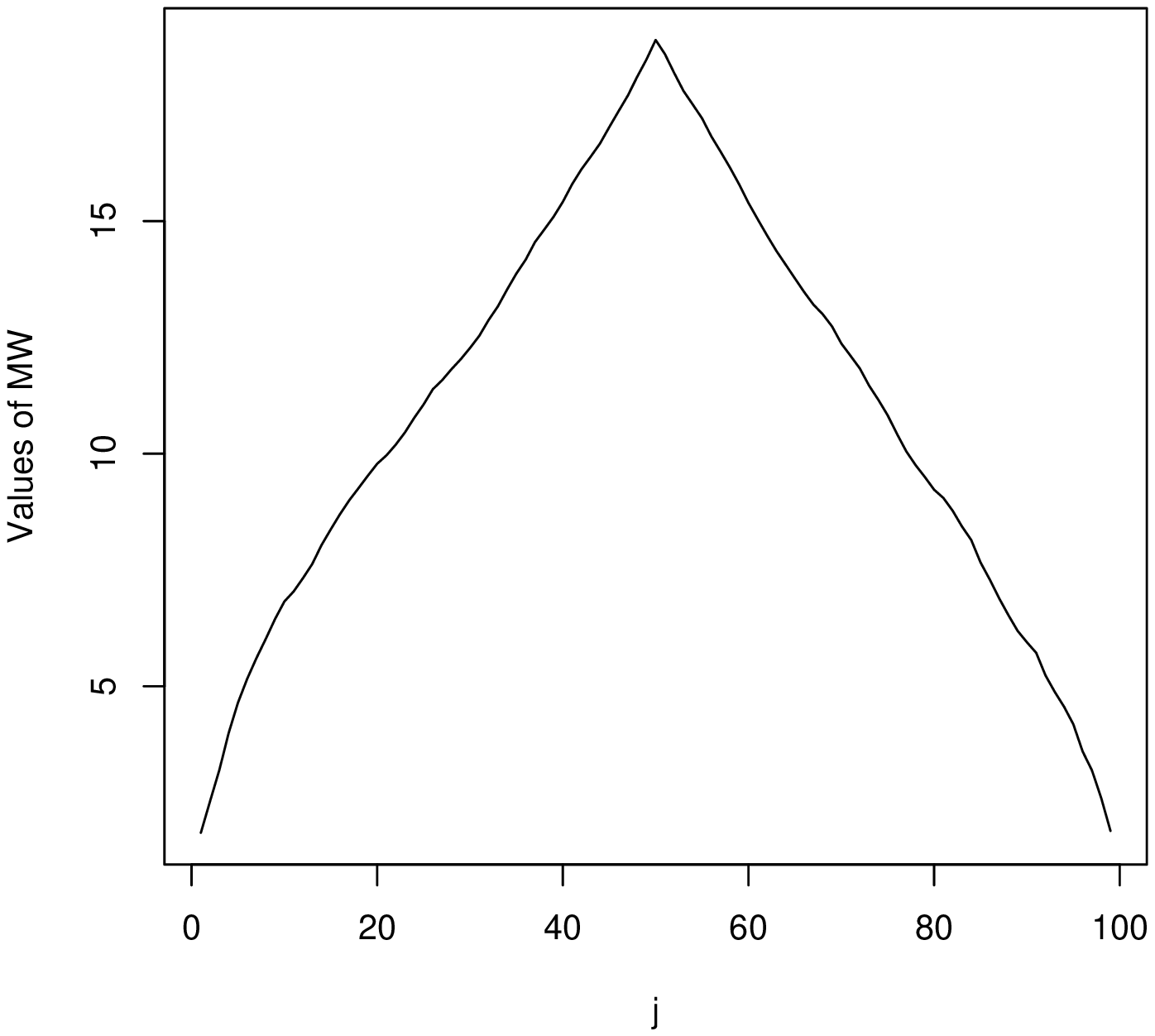}
\includegraphics[width=0.45\linewidth]{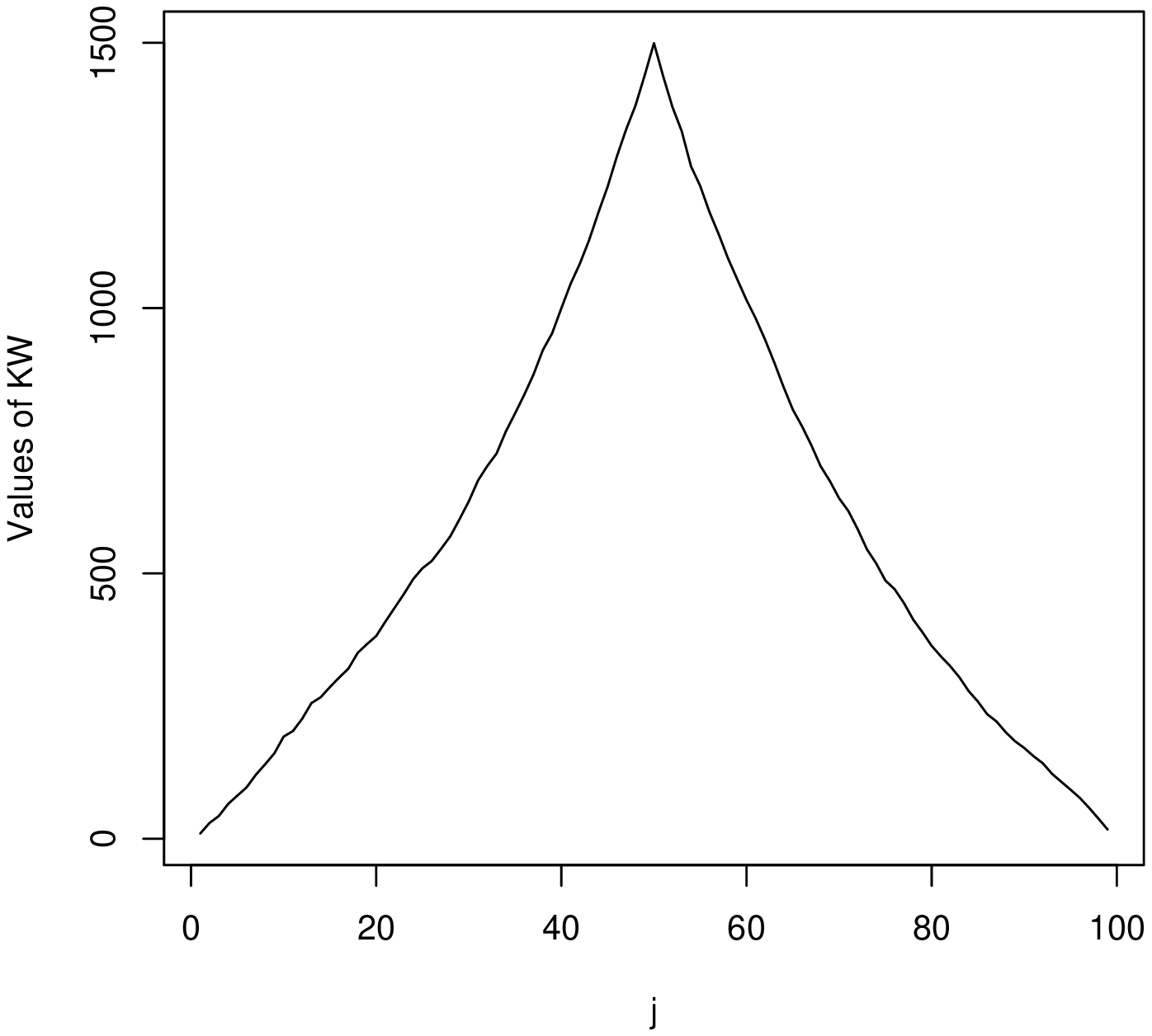}
\includegraphics[width=0.45\linewidth]{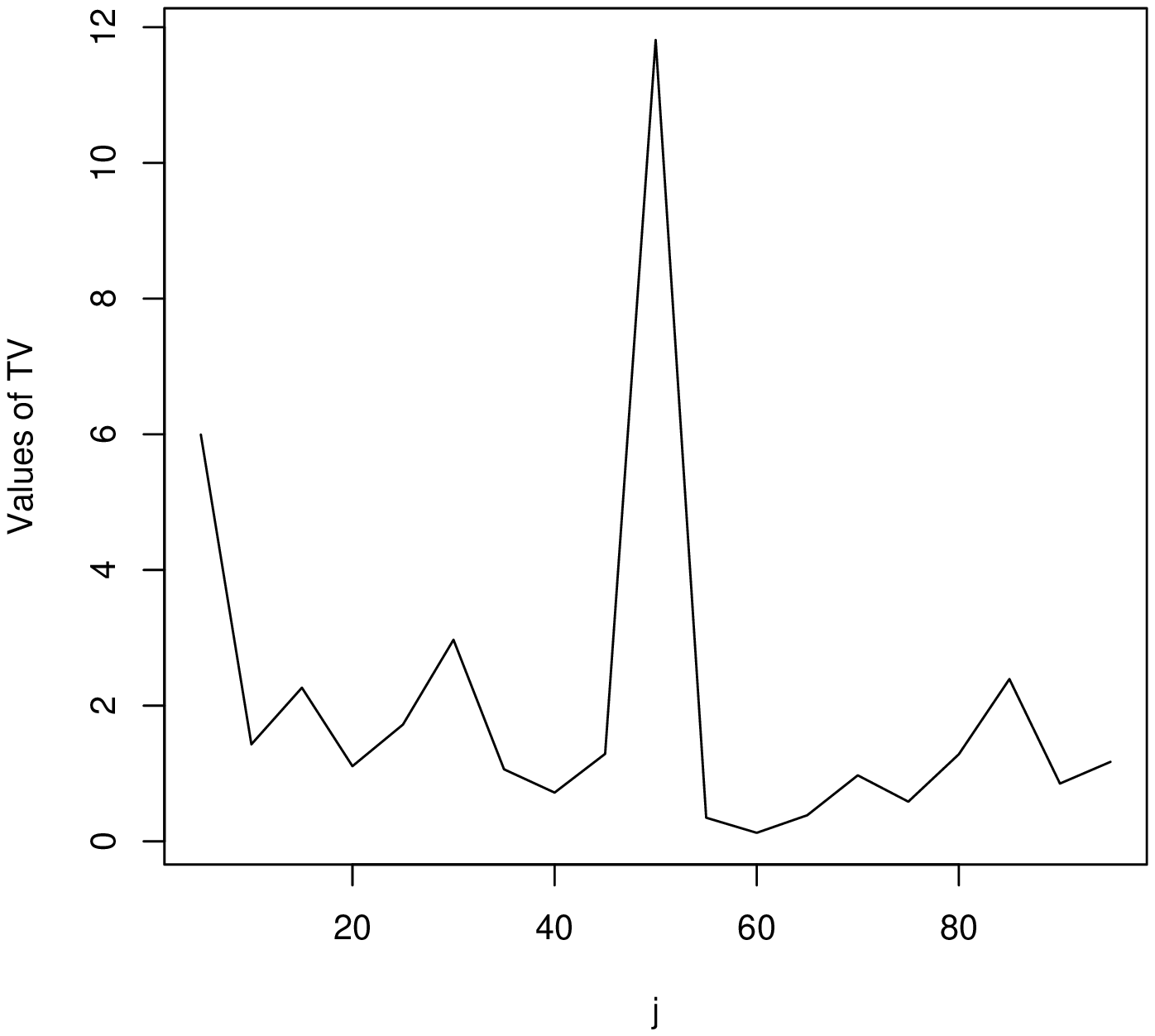}
\includegraphics[width=0.45\linewidth]{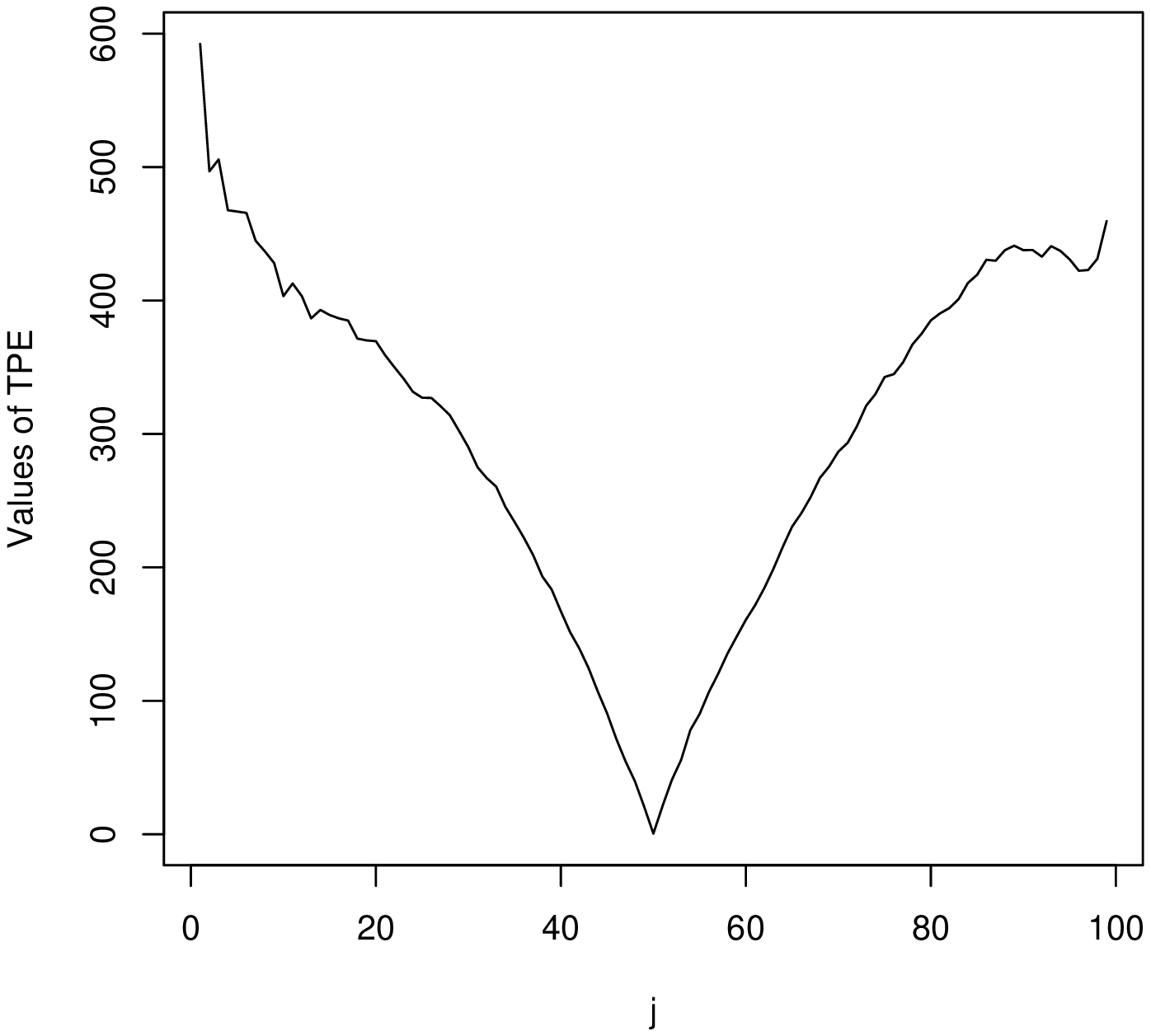}
\caption[Typical values of the test statistics]{Test statistic values: Mann-Whitney, Kruskal-Wallis, T.\ Variances and TPE}\label{funcion_mann}
\end{figure}

Figure~\ref{funcion_mann} shows values of the of the Mann-Whitney, Kruskal-Wallis, Squa\-red Ranks and the TPE Empirical test statistics taken along a straight line segment of simulated data.
The corresponding positions of the maxima (Mann-Whitney, Kruskal, Variance) or minimum (TPE) are taken to be the transition point between the two regions.
Data generation was carried in such a way that the true transition point is at $j=50$.
We note that all four methods successfuly identify the edge point. 

\section{Numerical results}\label{chap:results}

This section presents simulations performed to obtain a quantitative assessment of the performances of the proposed edge detection methods.
All simulations were run on personal computers with Intel\copyright\
Pentium\copyright\ IV CPUs of \unit[$3.20$]{GHz} running Windows~XP operating system.
The programming language used was \texttt{Ox} version $4.10$ \citep[for details, see][]{Doornik98}.
All graphics were produced using  \texttt{R} version~2.6.1 \citep{VenablesRipley:S:02}.

We shall evaluate the errors made when estimating the edge point (local error) in several parametric situations. 
For each situation, $1,000$ simulated rectangular windows of sizes $20 \times 100$ ($20$ rows, $100$ columns) are filled with with samples from $\mathcal{G}^0_{\mathcal{I}}$ distributions.

Each window is composed of two halves, and we consider all possible combinations of roughness parameters $-\alpha_\ell \in\{3, 4, 6, 8, 10, 12, 14, 16, 18, 20\}$ and $-\alpha_r \in \{2, 3, 4, 5, 6, 7, 8 , 9, 10, 11, 12, 13, 14, 15, 16, 17, 18,19, 20\}$, and number of looks $L \in \{1,3,8\}$. 
Here, $\alpha_\ell$ ($\alpha_r$) denotes the value of $\alpha$ used for data generation to the left (right) of the true edge point. 
These values span a variety of images often encountered in practice, ranging from single-loook to smoothed multi-look.
The scale parameter $\gamma$ equals
\begin{equation*}
\gamma_{\alpha,L}=\frac{\Gamma(-\alpha)\Gamma(L)L}{\Gamma(-\alpha-1)\Gamma(L+1)},
\end{equation*}
so that the distribution mean equals one in all situations.

It is noteworthy that edge detection algorithms are therefore evaluated in a very difficult situation, namely, when the areas have the same mean and only differ in texture.

Situations in which $\alpha_\ell=\alpha_r$ are not considered, since they entail no edge.
We thus consider $((10\times19)-10)\times3=540$ ($\alpha_\ell$, $\alpha_r$ and $L$) cases. For each of them, we simulate $1,000$ windows.
In each of these $540,000$ windows, the edge is detected by the five techniques already described, and the error is defined as the absolute difference between the true edge (which is located at position 50) and the detected edge.

The errors are stored in an $1000$-dimensional array $D_M(j)$ defined in the following way:
\begin{equation*}
D_M(j)=|50 - P_T(j)|, \quad j=1,2,\ldots, 1000,
\end{equation*}
where $P_T(j)$, $j = 1,\ldots, 1000$, is the transition point identified by method $M$ in the $j$th sample.
Note that $M$ ranges in the set
$$
\{\text{Gambini},\allowbreak \text{Kruskal},\text{Mann-Whitney},\text{Variance},\text{TPE}\}.
$$
Let $f(M)$ denote the percentage of times in which the estimated edge is more than $5$ pixels apart from the true value, using method $M$, i.e., 
\begin{equation*}
f(M)=\frac{\# \{ j\in \{1, \ldots , 1000\} \colon D_M(j)>5 \} }{1000}.
\end{equation*}
These are the error rates we report.
Method $M_i$ is considered to be more accurate than method $M_j$ whenever $f(M_i) < f(M_j)$, i.e., whenever the percentage of errors of method $M_i$ is smaller than that of method $M_j$, except for errors up to 5 pixels. 

\begin{figure}[hbt]
\centering
\subfigure[$L=1$, $\alpha_\ell=-8$\label{fig:1-3}]{\includegraphics[width=0.48\linewidth]{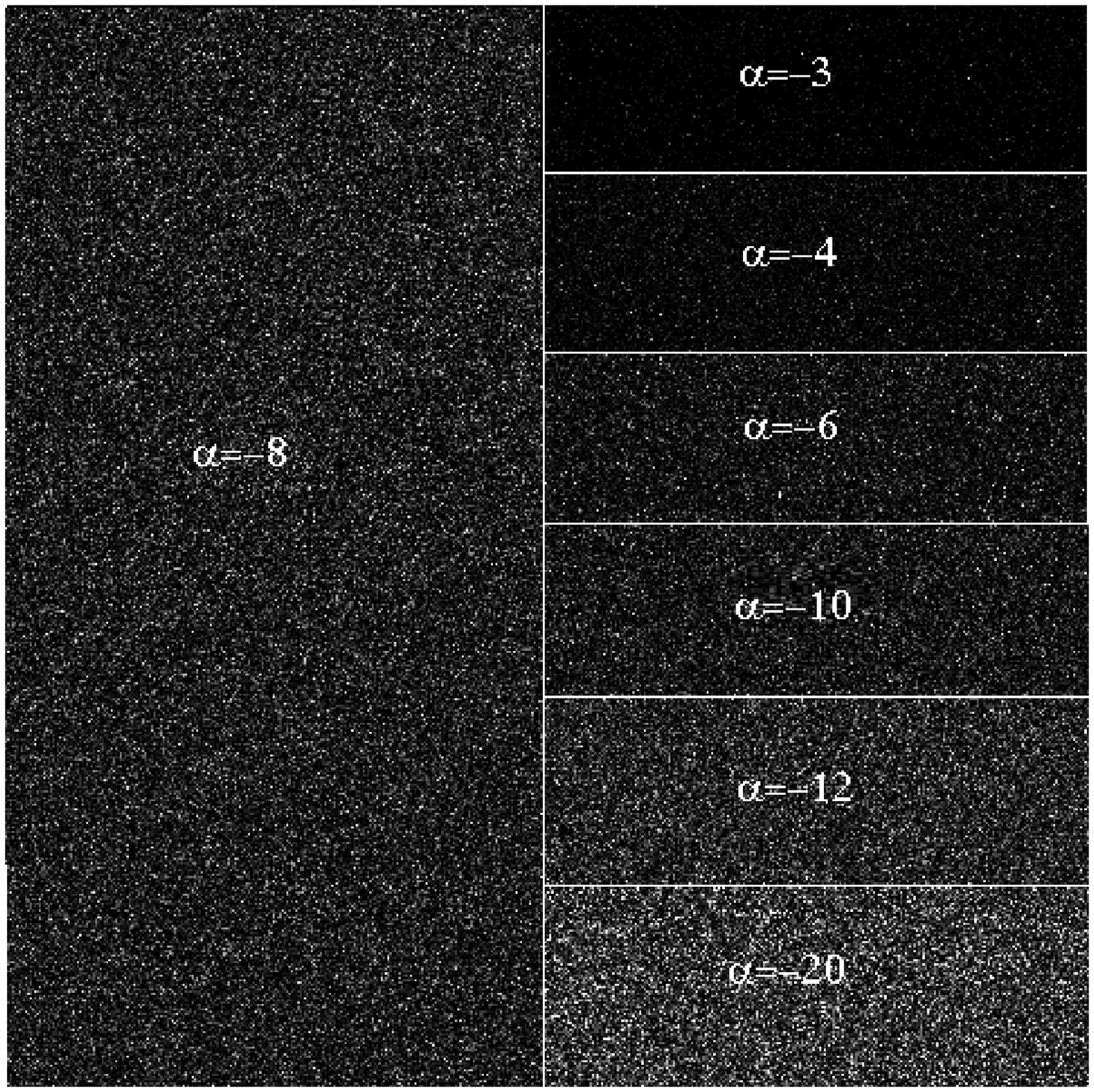}}
\subfigure[$L=1$, $\alpha_\ell=-12$\label{fig:1-8}]{\includegraphics[width=0.48\linewidth]{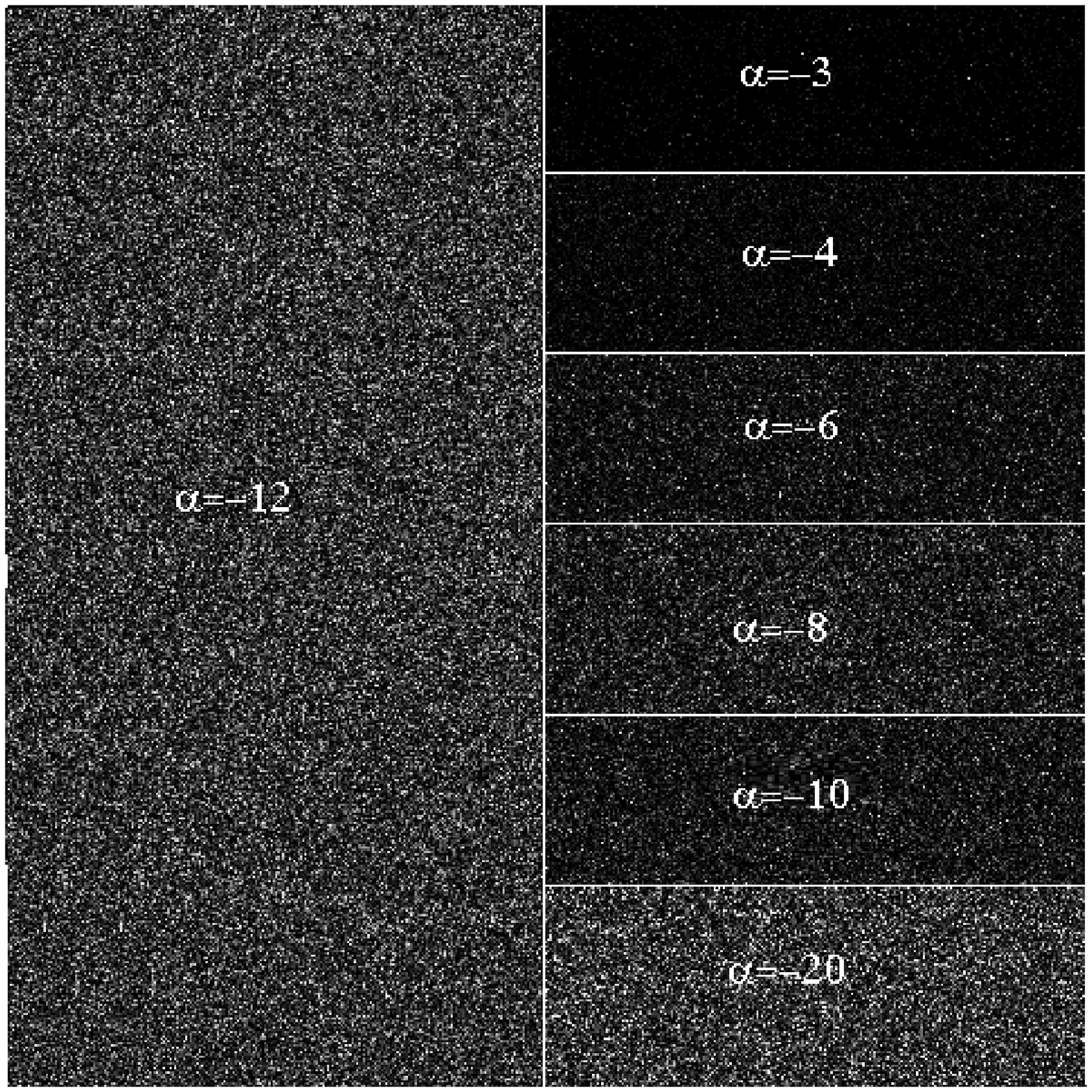}}\\
\subfigure[$L=1$, $\alpha_\ell=-18$\label{fig:1-12}]{\includegraphics[width=0.48\linewidth]{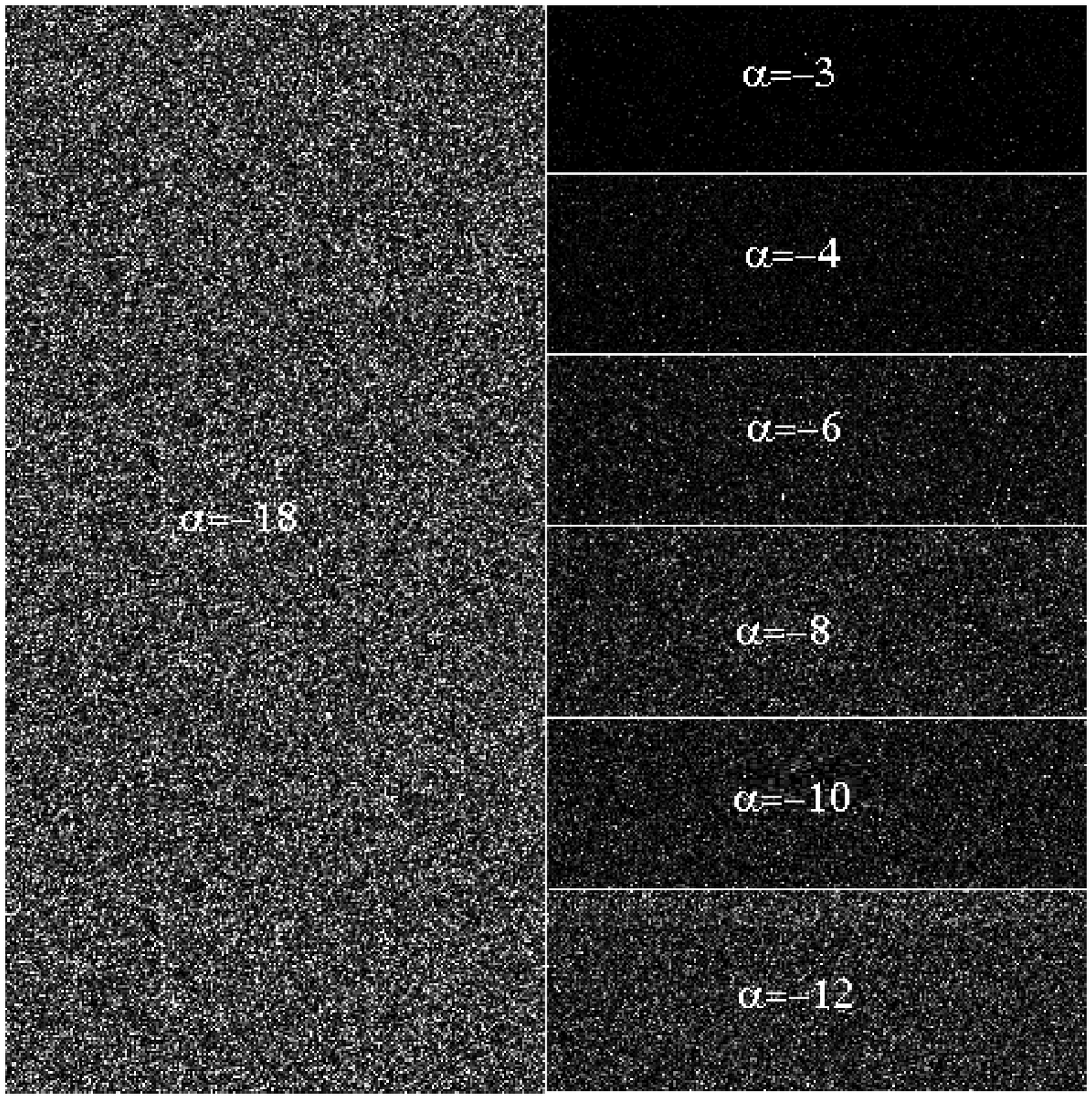}}
\subfigure[$L=8$, $\alpha_\ell=-8$\label{fig:8-18}]{\includegraphics[width=0.48\linewidth]{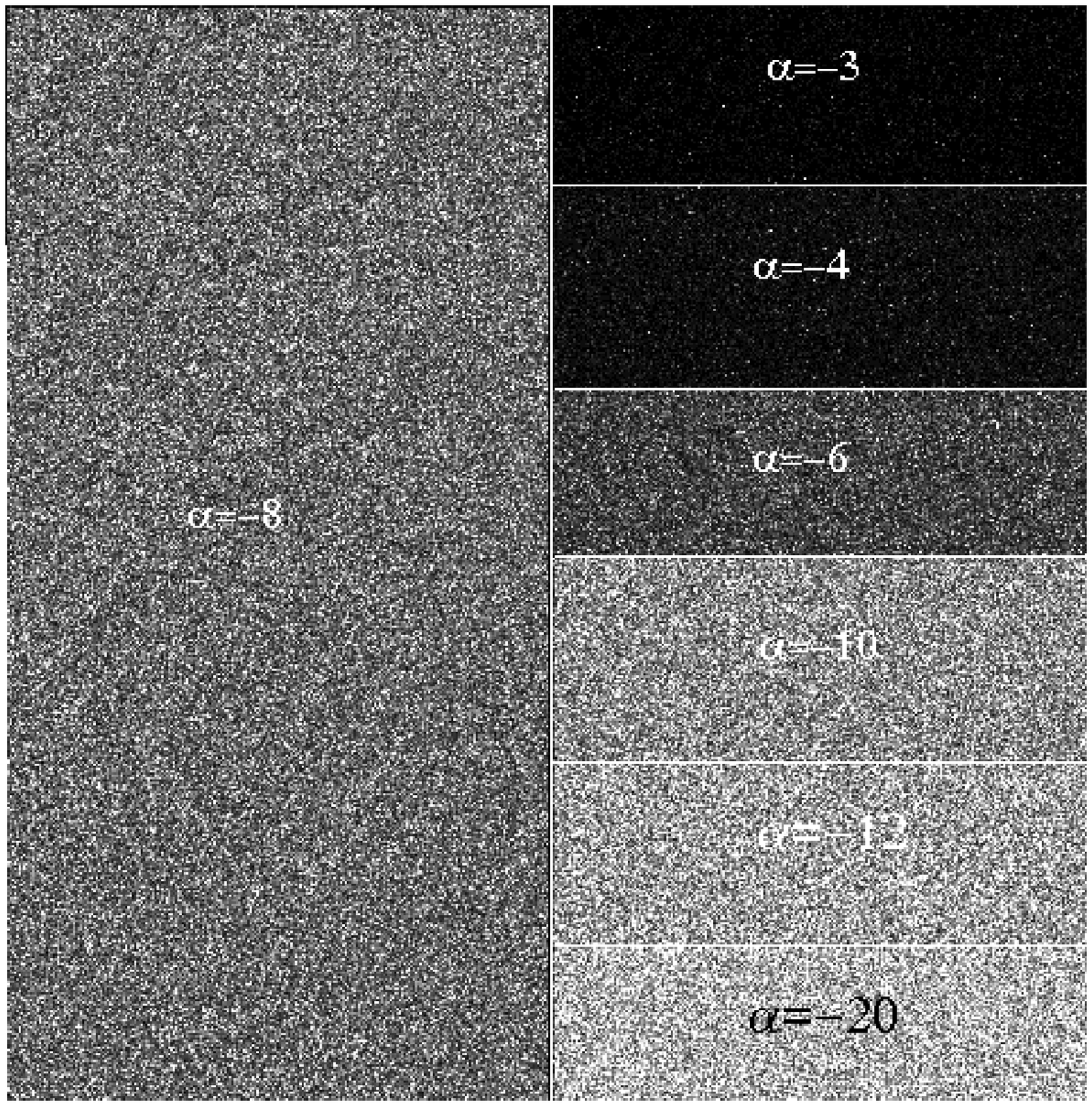}}
\caption{Four situations assessed, varying roughness to the left and strips $\alpha_r=\{-4,-6,-8,-10,-12,-20 \}$ to the right}\label{fig:foursituations}
\end{figure}

Figure~\ref{fig:foursituations} depicts four of the situations assessed: three where single ($L=1$) look data are contrasted (Figures~\ref{fig:1-3}, \ref{fig:1-8} and~\ref{fig:1-12}), and one with $L=8$ looks (Figure~\ref{fig:8-18}).
The single look data show the differences between heterogeneous ($\alpha_\ell=-8$, Figure~\ref{fig:1-3}), and homogeneous ($\alpha_\ell=-12$, Figure~\ref{fig:1-8}); $\alpha_\ell=-18$, Figure~\ref{fig:1-12}) areas and strips of varying roughness.
The $L=8$ data only presents the difference between a homogenous area ($\alpha_\ell = -18$) and several other strips (Figure~\ref{fig:8-18}).
The areas to the left are formed by independent draws from the $\mathcal G^0_{\mathcal I}(\alpha_\ell,\gamma_{\alpha_\ell,L}, L)$ distribution.
The strips to the right are formed by independent outcomes of the $\mathcal G^0_{\mathcal I}(\alpha_r,\gamma_{\alpha_r,L}, L)$ law with $\alpha_r=\{-4,-6,-8,-10,-12,-20 \}$.
The contrast among regions has been enhanced in order to make visualization easier; actual data are harder to differentiate.

\begin{figure}[hbt]
\centering
\subfigure[$\alpha_\ell=-8$, $L=1$\label{}]{\includegraphics[width=0.32\linewidth]{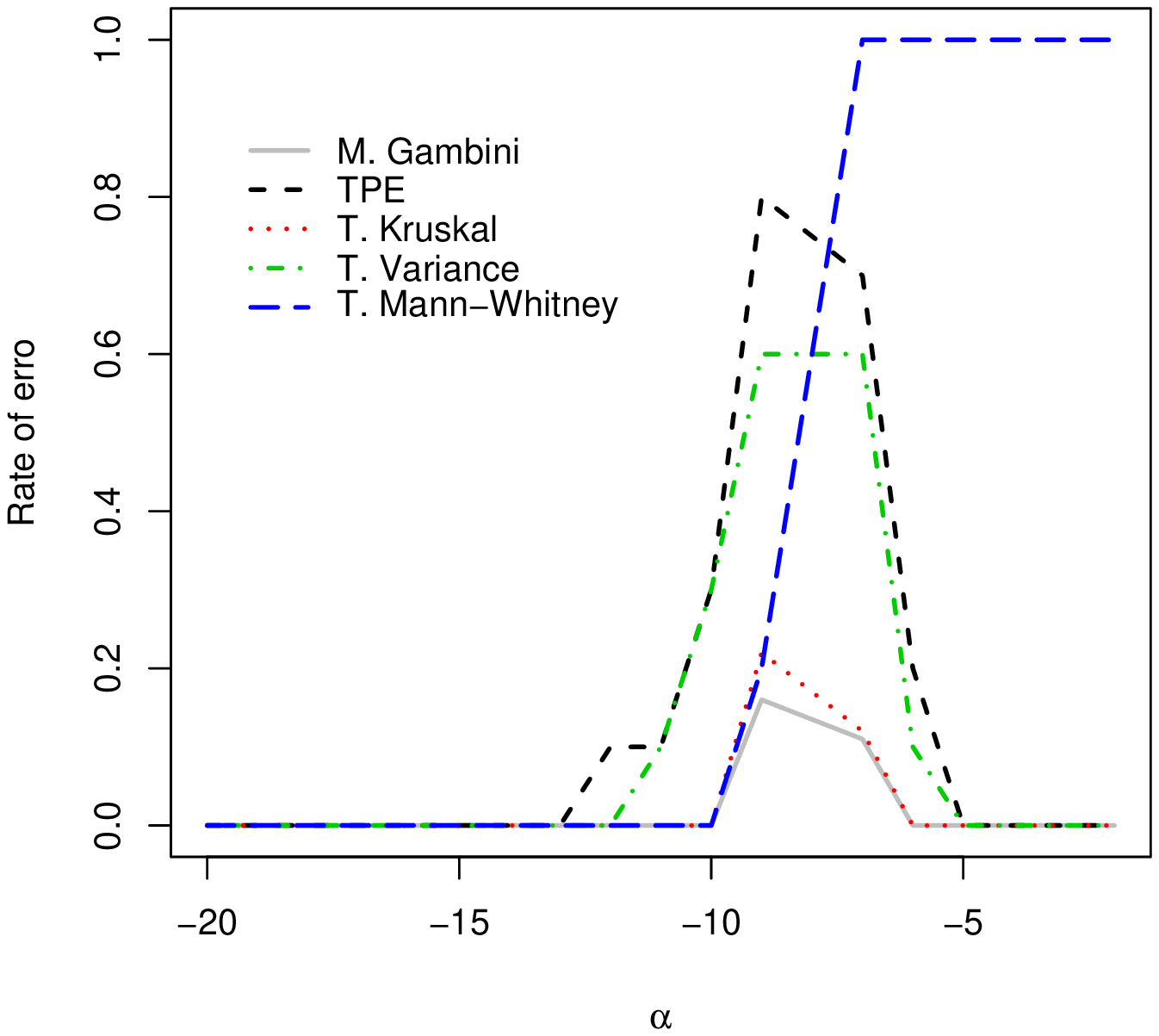}}
\subfigure[$\alpha_\ell=-8$, $L=3$\label{}]{\includegraphics[width=0.32\linewidth]{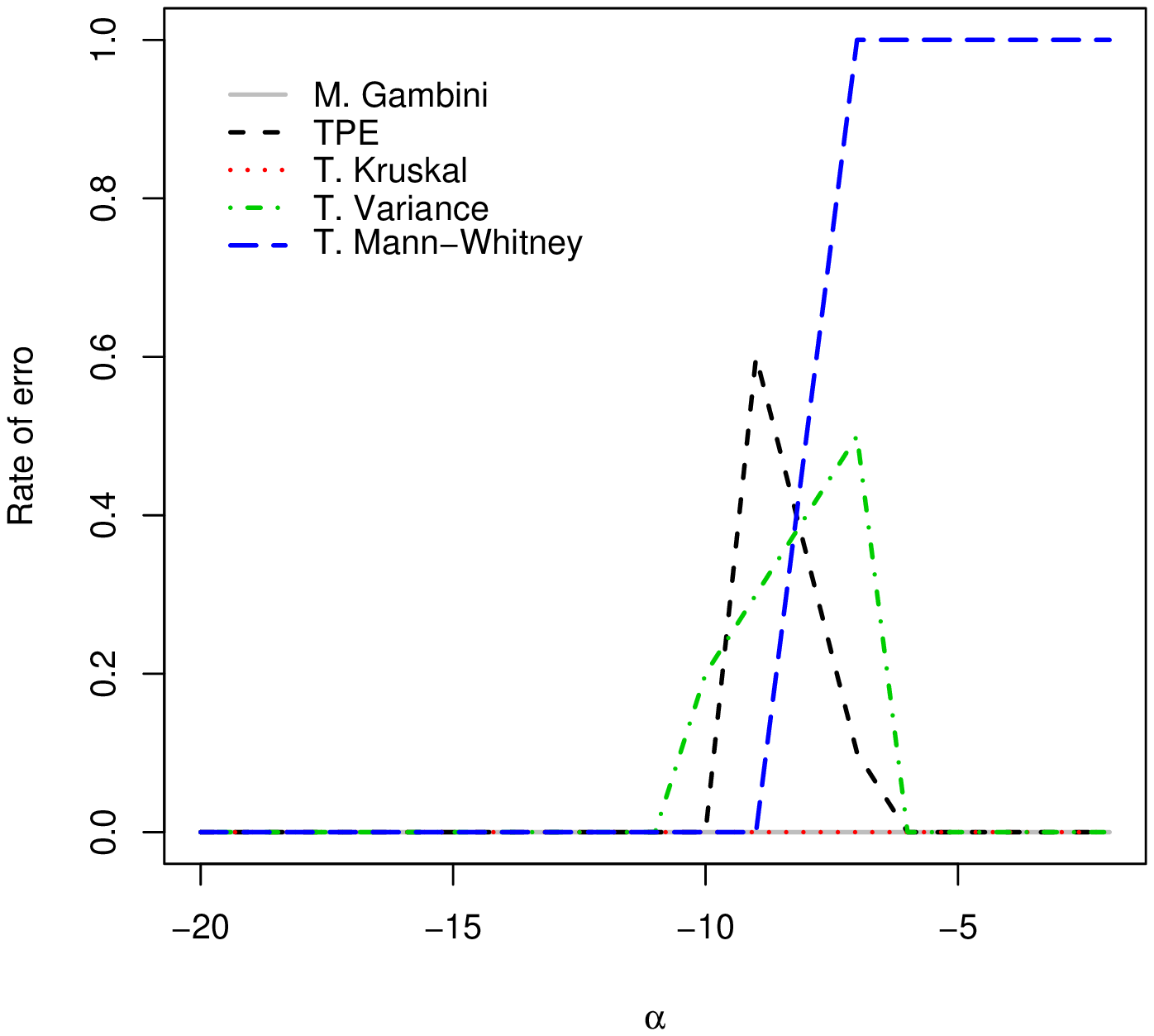}}
\subfigure[$\alpha_\ell=-8$, $L=8$\label{}]{\includegraphics[width=0.32\linewidth]{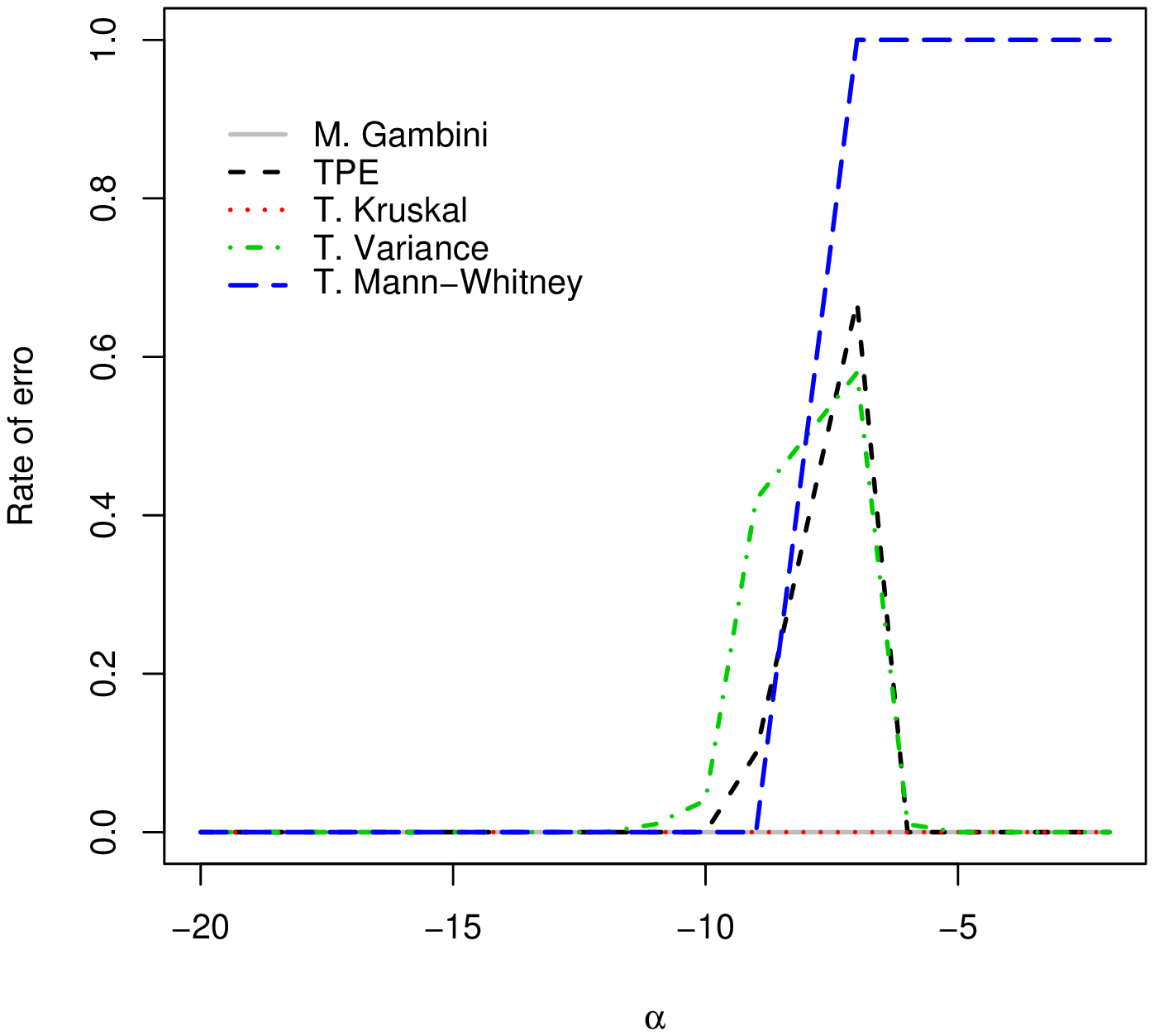}}\\
\subfigure[$\alpha_\ell=-12$, $L=1$\label{}]{\includegraphics[width=0.32\linewidth]{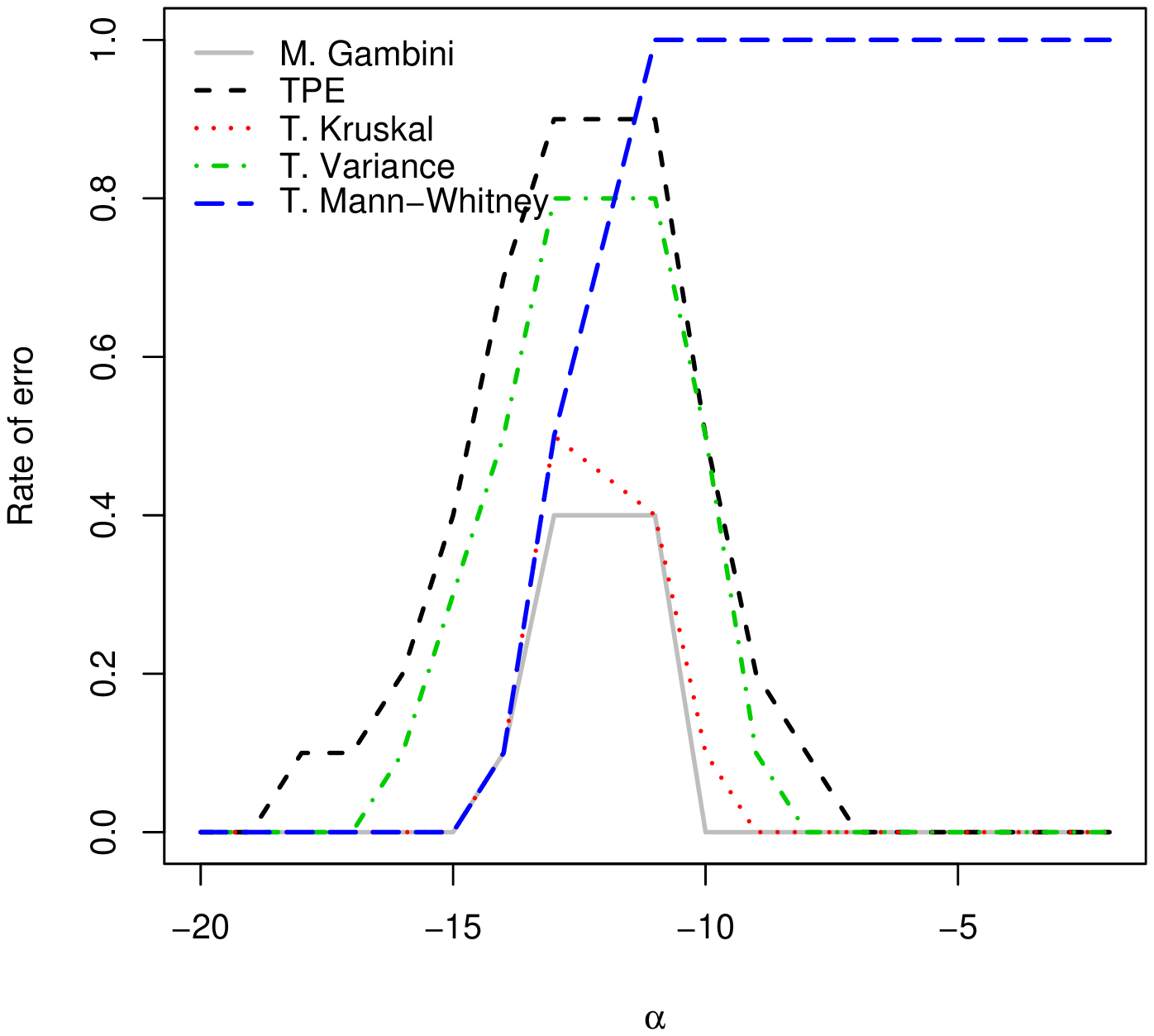}}
\subfigure[$\alpha_\ell=-12$, $L=3$\label{}]{\includegraphics[width=0.32\linewidth]{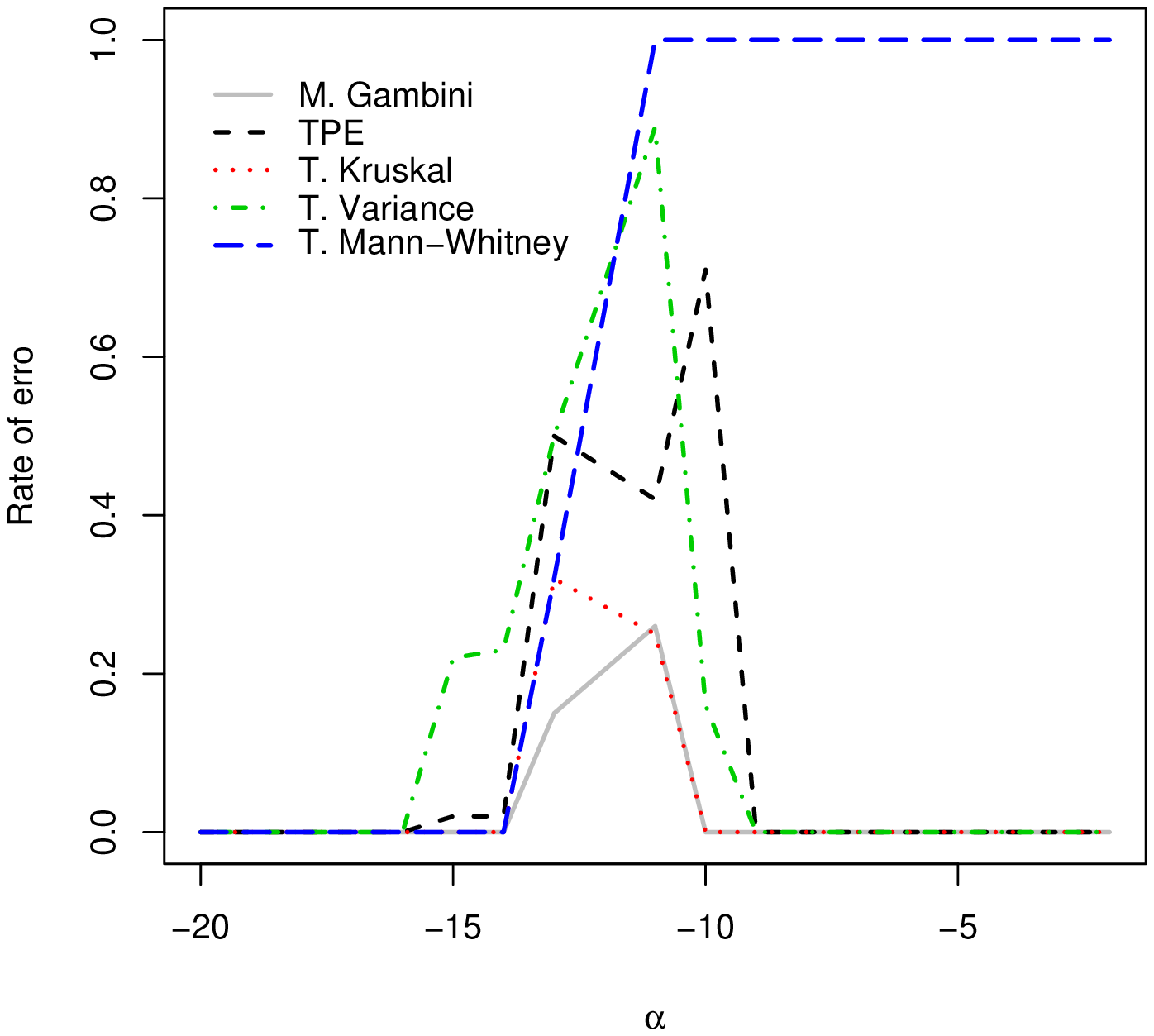}}
\subfigure[$\alpha_\ell=-12$, $L=8$\label{}]{\includegraphics[width=0.32\linewidth]{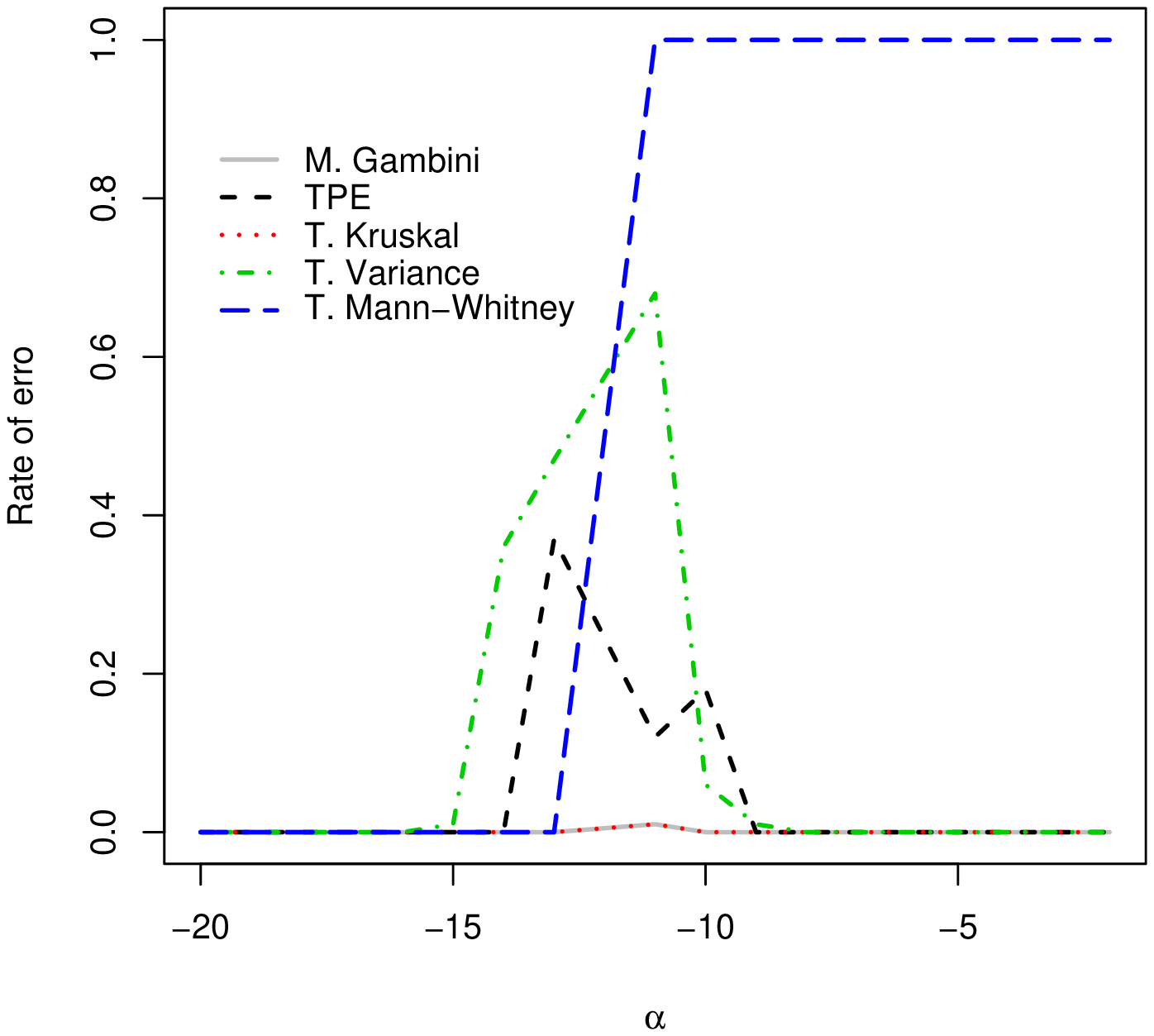}}\\
\subfigure[$\alpha_\ell=-18$, $L=1$\label{}]{\includegraphics[width=0.32\linewidth]{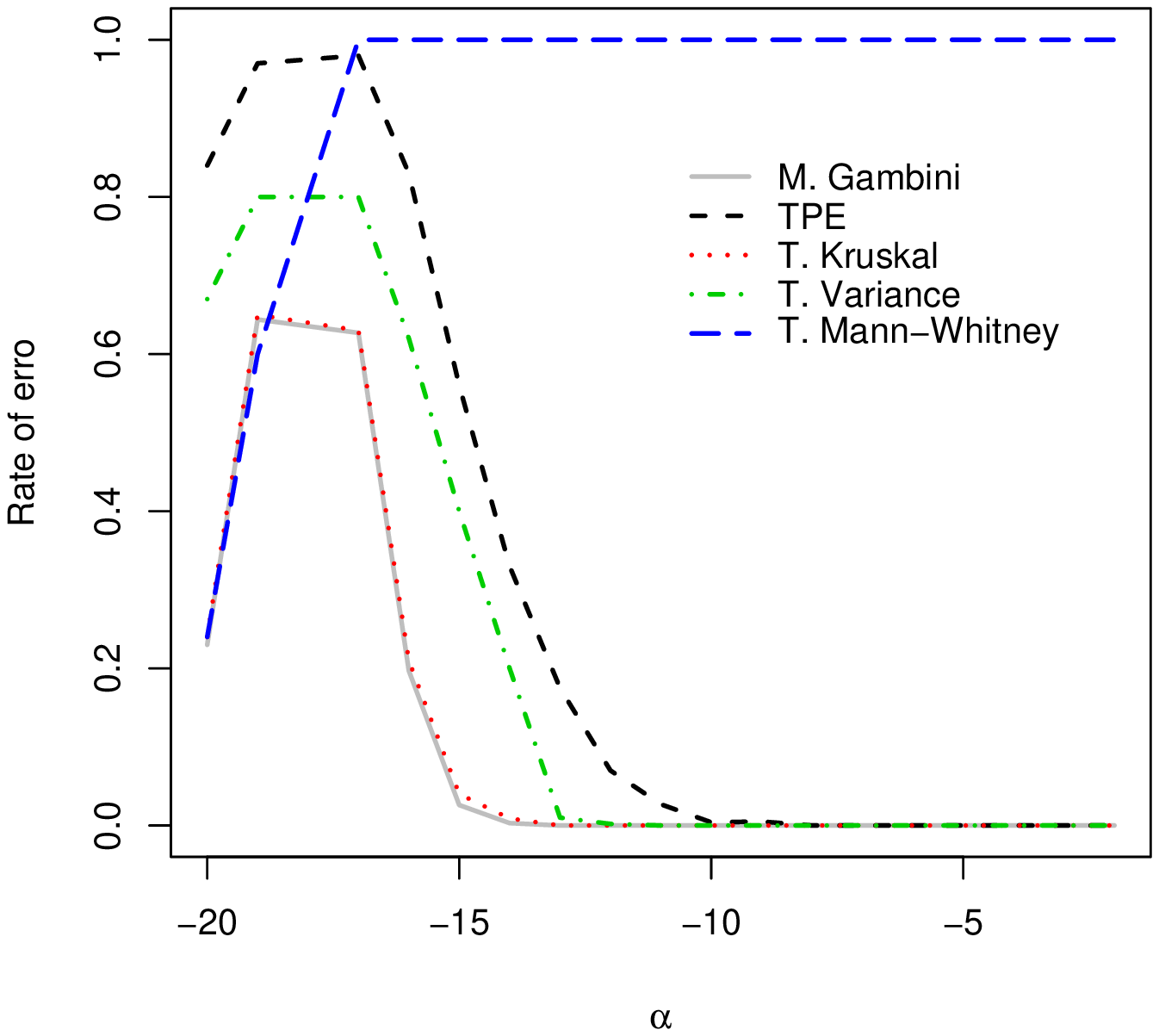}}
\subfigure[$\alpha_\ell=-18$, $L=3$\label{}]{\includegraphics[width=0.32\linewidth]{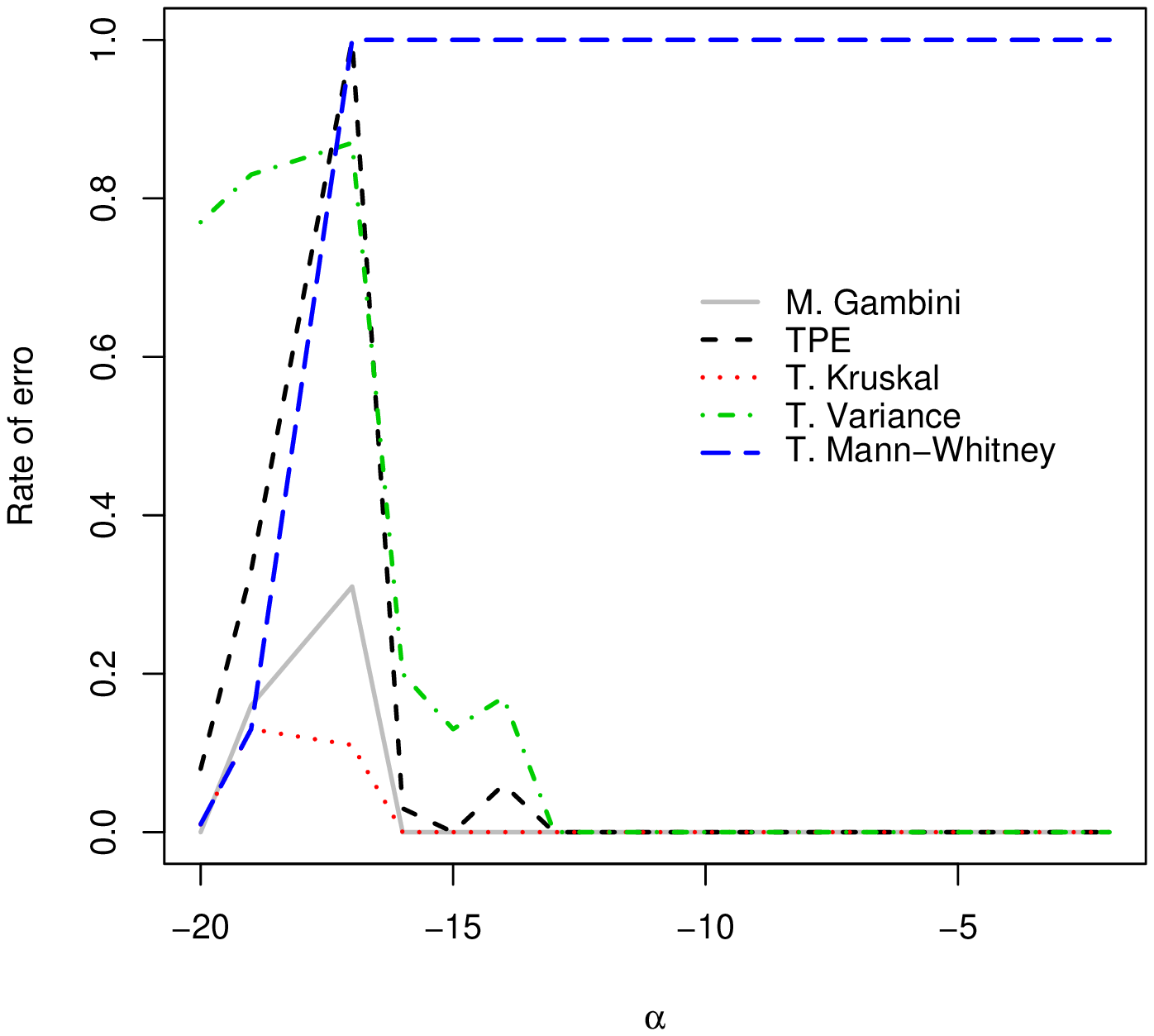}}
\subfigure[$\alpha_\ell=-18$, $L=8$\label{}]{\includegraphics[width=0.32\linewidth]{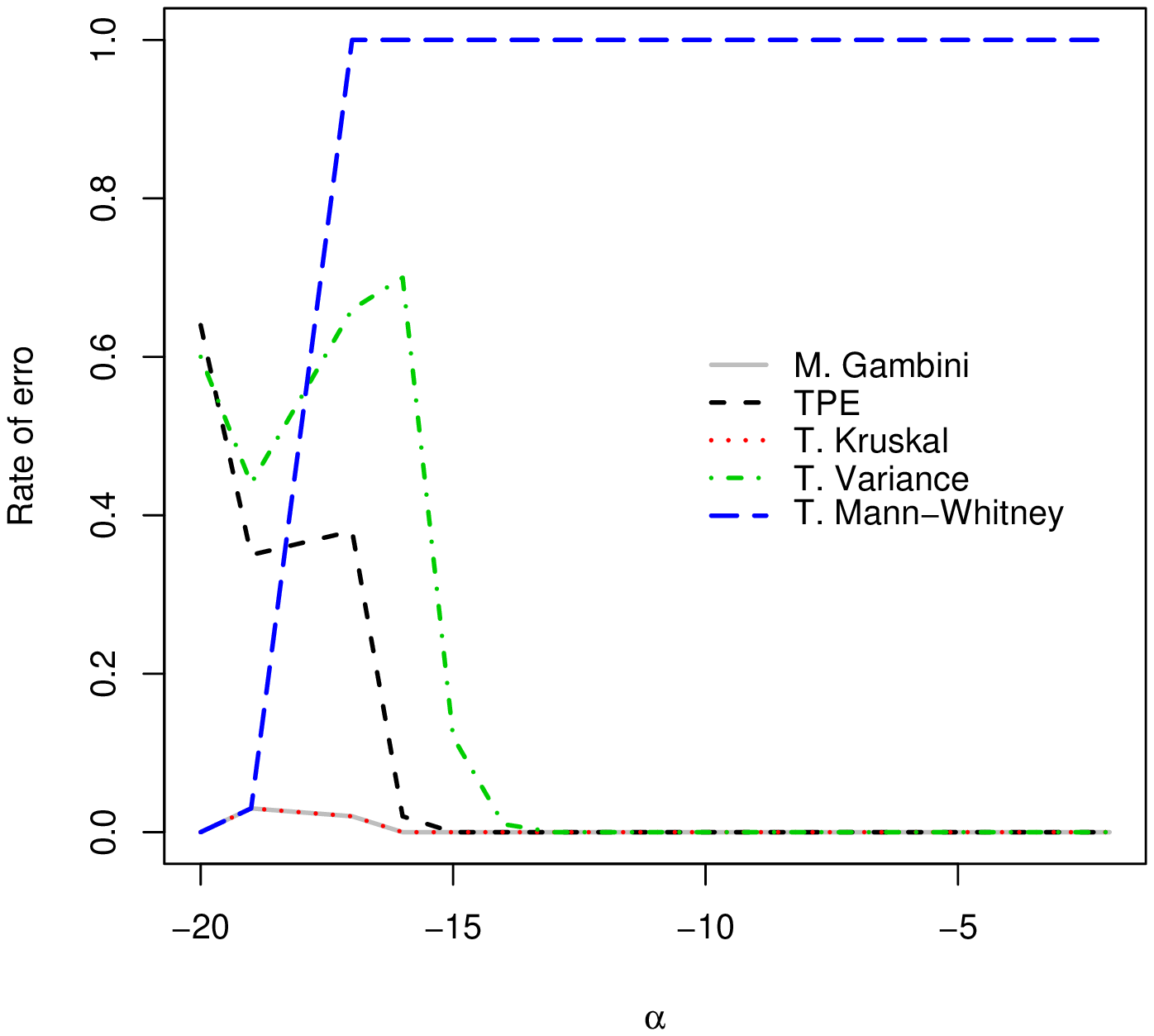}}
\caption{Error rates as functions of the number of looks ($L\in\{1,3,8\}$, left to right columns) and of the roughness ($\alpha_\ell\in\{-8,-12,-18\}$, top to bottom rows)}\label{fig:summaryerrorrates}
\end{figure}

Figure~\ref{fig:summaryerrorrates} summarizes the main results of our study, regarding error rates.
Each column presents the error rates as a function of $L$, the number of looks, provided a roughness for the left half of the strip $\alpha_\ell$.
We notice that the error rates are consistent with respect to $L$, since their behavior does not alter significantly.
Different lines, for a given $L$, show the behavior of the error rates as a function of the roughness.
\textbf{Mann-Whitney} is the only nonparametric technique that exhibits poor performance, but only when $\alpha_\ell>\alpha_r$; all other procedures are competitive.
\textbf{Kruskal} edge detection behaves similarly to \textbf{Gambini} procedure, with minor differences that do not consistently benefit either.
\textbf{TPE} and \textbf{Variance} are slightly worse than \textbf{Kruskal} and \textbf{Gambini}, mainly when multilook ($L=3,8$) imagery is used.
The error rates are consistently larger when the left and right roughnesses are similar; notice that the curves peaks shift to the left from top to bottom.
Overall, the \textbf{Kruskal} and \textbf{Gambini} error rates decrease as the number of looks increases, the same pattern holding for the degree of heterogeneity. 

\begin{figure}[hbt]
\centering
\subfigure[All techniques\label{fig:timesall}]{\includegraphics[width=0.45\linewidth]{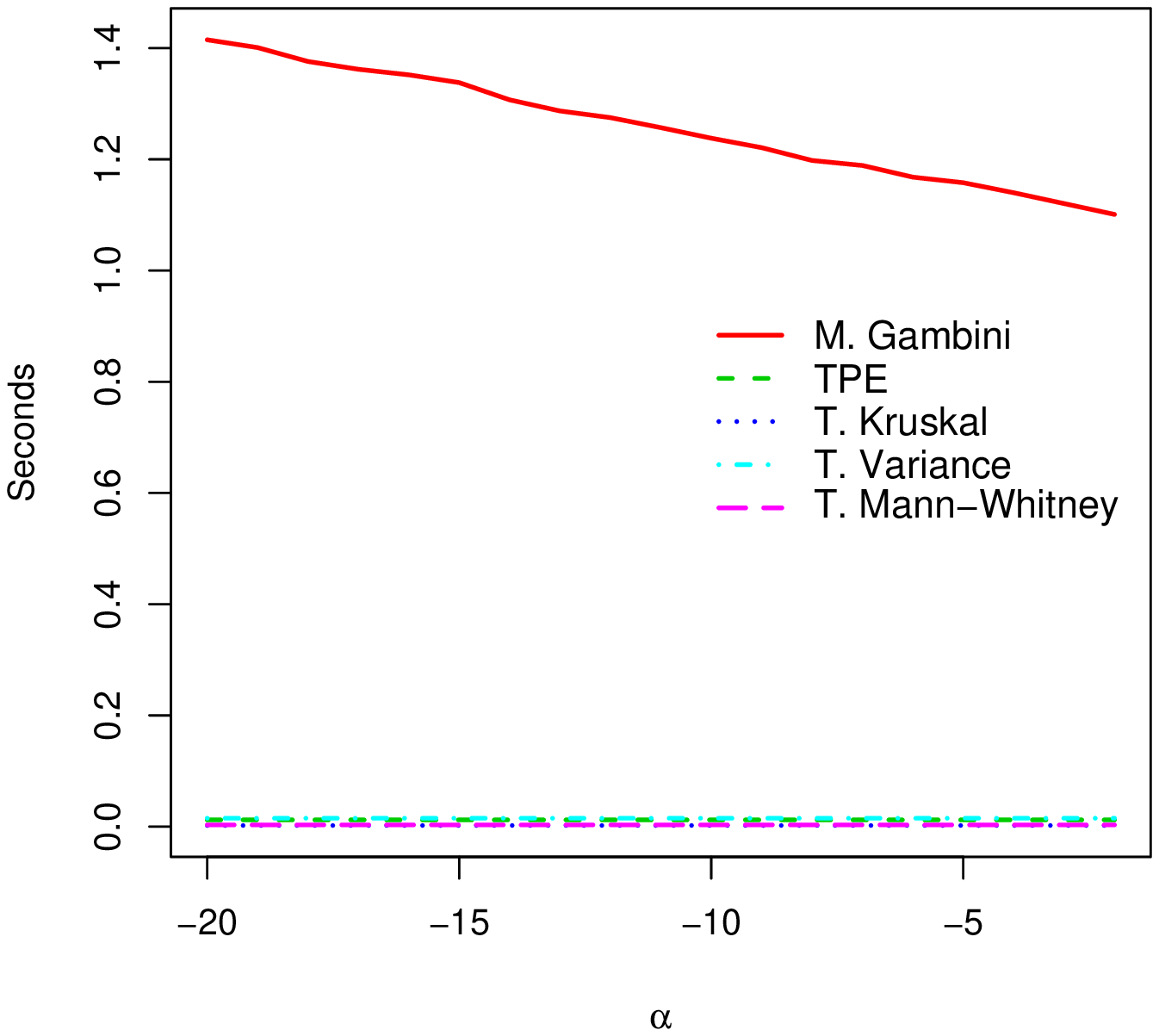}}
\subfigure[Nonparametric techniques\label{fig:timesnonparam}]{\includegraphics[width=0.45\linewidth]{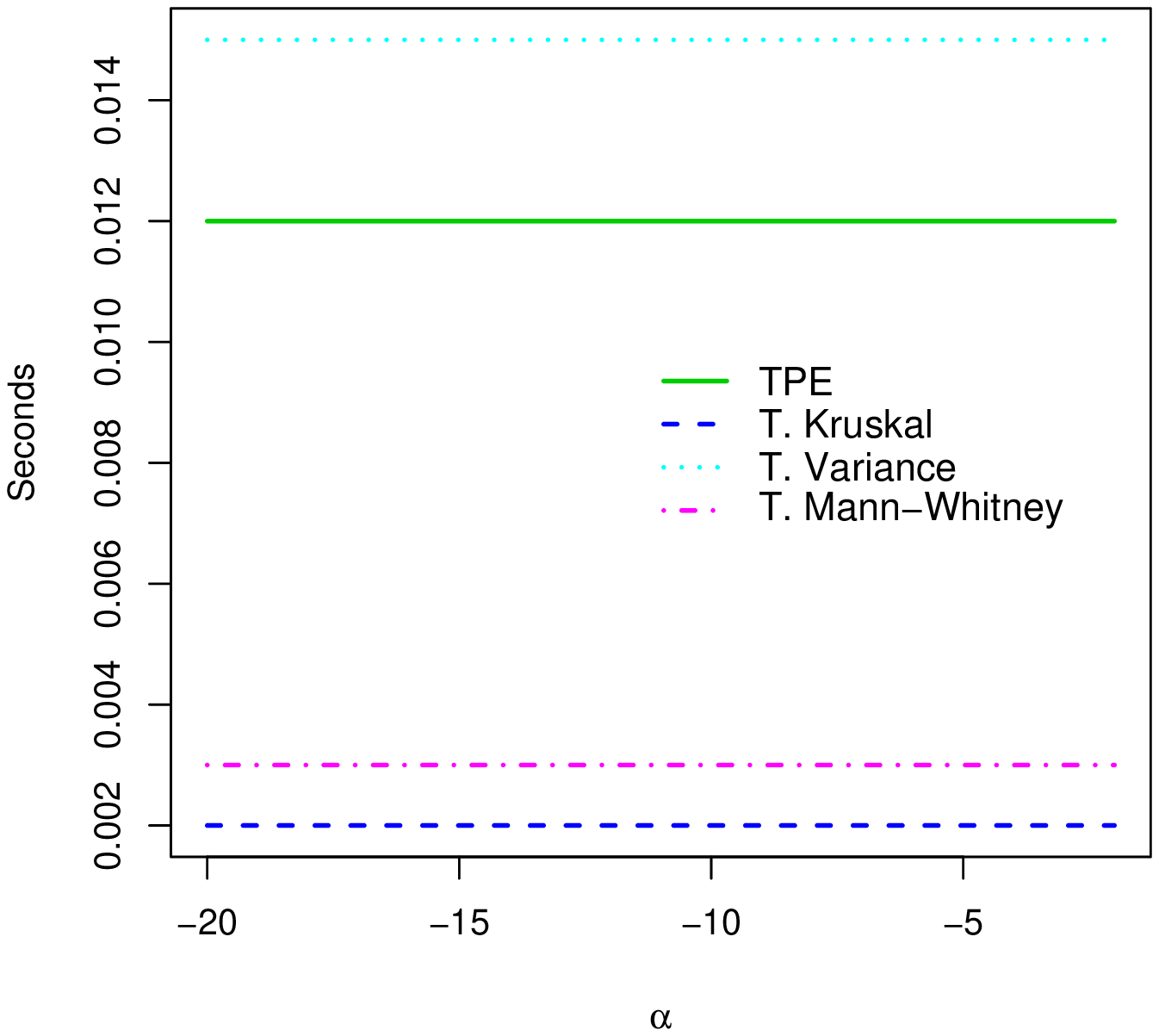}}
\caption{Average execution times (in seconds) for $\alpha_\ell=-3$ with $L=1$}\label{tiempos_a3_n1}
\end{figure}

Figure~\ref{tiempos_a3_n1} presents the average execution times for the $\alpha_\ell=-3$ and $L=1$ situation, which is representative of all remaining execution times.
As Figure~\ref{fig:timesall} shows, the time spent by \textbf{Gambini} decreases as the value of $\alpha_r$ increases, but it is consistently larger than the  nonparametric techniques detection times.
The differences in execution times are of three orders of magnitude and occur in all of the situations herein assessed.
Figure~\ref{fig:timesnonparam} shows the average execution times of the nonparametric methods, which never exceeded one second; \textbf{Kruskal} was the fastest followed by \textbf{Mann-Whitney}.

As noted earlier, the hardest task one can face is the discrimination of regions indexed by similar parameter values.
Table~\ref{tab:tough} presents a subset of our numerical results and focuses on the most challenging cases.
It is clear from the figures in this table that \textbf{Kruskal} is consistently faster than \textbf{Gambini}, reaching a scale of $10^3$ in some cases. The error rates of the two methods are similar.

\begin{table}[hbt]
\centering
\caption{Error rates and execution times of challenging situations}\label{tab:tough}
\begin{tabular}{rrrrrrr} \toprule
&& & \multicolumn{2}{c}{\textbf{Gambini}} & \multicolumn{2}{c}{\textbf{Kruskal}} \\ \cmidrule(r){4-5} \cmidrule(r){6-7}
\multicolumn{3}{c}{Situation} & Error & Time & Error & Time \\ \midrule
$L=1$ & $\alpha_\ell=-3$ & $\alpha_r=-2$ & $0.00$   & $1.101$  & $0.00$   & $0.002$ \\
      & 		 & $\alpha_r=-4$ & $0.10$   & $1.140$  & $0.00$   & $0.002$ \\
      & $\alpha_\ell=-8$ & $\alpha_r=-7$ & $11.20$  & $1.353$  & $12.20$  & $0.002$\\
      & 		 & $\alpha_r=-9$ & $16.20$  & $1.389$  & $22.40$  & $0.002$ \\
      & $\alpha_\ell=-12$& $\alpha_r=-11$ & $38.30$  & $1.544$  & $40.70$  & $0.002$ \\
      & 		 & $\alpha_r=-13$ & $40.30$  & $1.622$  & $45.90$  & $0.002$ \\
      & $\alpha_\ell=-18$& $\alpha_r=-17$ & $62.70$  & $1.893$  & $63.60$  & $0.002$ \\
      & 		 & $\alpha_r=-19$ & $64.40$  & $2.013$  & $65.00$  & $0.002$ \\ \midrule
$L=3$ & $\alpha_\ell=-3$ & $\alpha_r=-2$ & $0.00$  & $1.090$  & $0.00$  & $0.002$ \\
      & 		 & $\alpha_r=-4$ & $0.00$  & $1.117$  & $0.00$  & $0.002$ \\
      & $\alpha_\ell=-8$ & $\alpha_r=-7$ & $0.00$  & $1.269$  & $0.00$  & $0.002$ \\
      & 		 & $\alpha_r=-9$ & $0.00$  & $1.279$  & $0.00$  & $0.002$ \\
      & $\alpha_\ell=-12$& $\alpha_r=-11$ & $26.40$  & $1.491$  & $24.50$  & $0.002$ \\
      & 		 & $\alpha_r=-13$ & $14.90$  & $1.430$  & $31.50$  & $0.002$ \\
      & $\alpha_\ell=-18$& $\alpha_r=-17$ & $31.30$  & $1.324$  & $11.20$  & $0.001$ \\
      & 		 & $\alpha_r=-19$ & $15.70$  & $1.417$  & $13.10$  & $0.001$ \\ \midrule
$L=8$ & $\alpha_\ell=-3$ & $\alpha_r=-2$ & $0.00$  & $0.953$  & $0.00$  & $0.001$ \\
      & 		 & $\alpha_r=-4$ & $0.00$  & $0.949$  & $0.00$  & $0.001$ \\
      & $\alpha_\ell=-8$ & $\alpha_r=-7$ & $0.20$  & $1.160$  & $0.00$  & $0.001$ \\
      & 		 & $\alpha_r=-9$ & $0.00$  & $1.091$  & $0.00$  & $0.001$ \\
      & $\alpha_\ell=-12$& $\alpha_r=-11$ & $1.20$  & $1.326$  & $1.30$  & $0.001$ \\
      & 		 & $\alpha_r=-13$ & $0.30$  & $1.303$  & $0.40$  & $0.001$ \\
      & $\alpha_\ell=-18$& $\alpha_r=-17$ & $1.70$  & $1.423$  & $1.50$  & $0.001$ \\
      & 		 & $\alpha_r=-19$ & $3.10$  & $1.478$  & $3.40$  & $0.002$ \\ \bottomrule
\end{tabular}
\end{table}

\section{Application to real data}\label{chap:application}

\begin{figure}[hbt]
\centering
 \includegraphics[width=.7\linewidth]{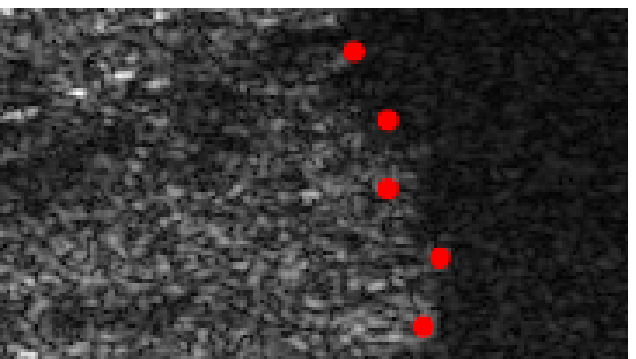}
\caption{ESAR image and detected edge points}\label{fig:application}
\end{figure}

Figure~\ref{fig:application} presents a SAR image for which the estimated number of looks is $3.2$.
The window considered has $101\times181$ pixels, and was obtained over agricultural fields of Oberpfaffenhofen, Germany, in the L-band, by the ESAR sensor~\cite{SupervisedClassificationConditionalRandomFieldsMultiscaleRegionConnectionCalculusSAR}.

The original data displayed almost no differences, and it was enhanced for visual purposes.
Two main regions can be observed, namely: the dark (left) and light (right) areas.
The estimated roughness parameter in both areas are approximately equal to 7.5 (i.e., $\widehat{\alpha}=7.5$), thus implying that both areas are slightly heterogeneous, probably due to relief.

An edge was detected in each of five non-overlapping strips of $20\times181$ pixels.
The Mann-Whitney, Kruskal-Wallis and Gambini estimates agree and are indicated by red dots in the middle of each strip.
The estimates, thus, provide accurate starting points for any subsequent edge detection algorithm.

\section{Conclusions}\label{chap:conclu}

Our chief goal was to propose alternative techniques for edge detection in speckled imagery. The proposed methods were compared to that of \citet{Gambini:StatisticsComputing,Gambini2006}.
The techniques here assessed do not try to eliminate existing speckle, but to extract information from its statistical properties.
The model used to describe these data is the $\mathcal{G}^0_{\mathcal{I}}$ distribution, which \citet{Mejail2001,Mejail2003} show can be used as an universal model.

The methods under assessment aim at identifying edges between regions with different degrees of roughness, which, in turn, is determined by $\alpha$, the roughness parameter of the $\mathcal{G}^0_{\mathcal{I}}(\alpha,\gamma,n)$ distribution.
Homogenous, e.g., pastures, heterogeneous regions, e.g., forest, and very heterogeneous
regions, e.g., urban, targets are considered.

In order to compare the performances of the five methods, a Monte Carlo experiment was carried out.
Two criteria were employed in the comparison, namely: the error made in detecting an edge and the execution time.
The thoughest setting was considered, namely, when the areas on both sides of the edge have the same mean and number of looks, differing only on the roughness.

From the experimental results it was observed that the \textbf{Kruskal} detector performed slightly better than \textbf{Gambini}.
Compared with the \textbf{TPE}, \textbf{Variance} and \textbf{Mann-Whitney} detectors, the \textbf{Kruskal} performed much better.

It is noteworthy that all detection methods perform well when the edge separates areas with very different degress of roughness.

When the generated image represents homogenous zones, the error frequencies of all methods are relatively high, nevertheless this problem is alleviated by increasing the number of looks $L$, causing the edge detection to become more accurate.

The \textbf{Kruskal} detector displays the best results, both with respect to error and execution time.
The latter is, in many cases, $1000$ times smaller than that of \textbf{Gambini}.

The \textbf{TPE} method for edge detection performs well when $\alpha_\ell$ differs from $\alpha_r$ by at least three units.

\textbf{Mann-Whitney} has good performance (both in terms of error rate and speed) when $\alpha_\ell < \alpha_r$.
When $\alpha_\ell > \alpha_r$ the method is unable to locate the edge, mostly because it only accounts for the ranks of one of the samples.

Overall, our results suggest that the \textbf{Gambini} edge detection technique can be successfully replaced by our \textbf{Kruskal}. By doing so, one achieves similar precision at a much lower computational cost (the latter is approximately one thousand times faster).

A promising line of research is the use of stochastic distances, as derived by \citet{NascimentoCintraFrertIEEETGARS}.

\section*{Acknowledgements}

The authors gratefully acknowledge research grants from Capes and CNPq. We also thank an anonymous referee for comments and suggestions. 

\bibliographystyle{model1b-num-names}
\bibliography{edwin}

\begin{thebibliography}{27}
\expandafter\ifx\csname natexlab\endcsname\relax\def\natexlab#1{#1}\fi
\providecommand{\bibinfo}[2]{#2}
\ifx\xfnm\relax \def\xfnm[#1]{\unskip,\space#1}\fi
\bibitem[{Allende et~al.(2006)Allende, Frery, Galbiati and
  Pizarro}]{Allende2006}
\bibinfo{author}{H.~Allende}, \bibinfo{author}{A.C. Frery},
  \bibinfo{author}{J.~Galbiati}, \bibinfo{author}{L.~Pizarro},
  \bibinfo{title}{M-estimators with asymmetric influence functions: the
  $\mathcal{G}^0_{A}$ distribution case}, \bibinfo{journal}{Journal of
  Statistical Computation and Simulation} \bibinfo{volume}{76}
  (\bibinfo{year}{2006}) \bibinfo{pages}{941--956}.
\bibitem[{Beauchemin et~al.(1998)Beauchemin, Thomson and
  Edwards}]{Beauchemin1998}
\bibinfo{author}{M.~Beauchemin}, \bibinfo{author}{K.P.B. Thomson},
  \bibinfo{author}{G.~Edwards}, \bibinfo{title}{On nonparametric edge detection
  in multilook {SAR} images}, \bibinfo{journal}{IEEE Transactions on Geoscience
  and Remote Sensing} \bibinfo{volume}{36} (\bibinfo{year}{1998})
  \bibinfo{pages}{1826--1829}.
\bibitem[{Bovik et~al.(1986)Bovik, Huang and Munson}]{BOVIK1986}
\bibinfo{author}{A.C. Bovik}, \bibinfo{author}{T.S. Huang},
  \bibinfo{author}{D.C. Munson}, \bibinfo{title}{Nonparametric-tests for
  edge-detection in noise}, \bibinfo{journal}{Pattern Recognition}
  \bibinfo{volume}{19} (\bibinfo{year}{1986}) \bibinfo{pages}{209--219}.
\bibitem[{Brigger et~al.(2000)Brigger, Hoeg and Unser}]{Brigger2000}
\bibinfo{author}{P.~Brigger}, \bibinfo{author}{J.~Hoeg},
  \bibinfo{author}{M.~Unser}, \bibinfo{title}{B-spline snakes: A flexible tool
  for parametric contour detection}, \bibinfo{journal}{IEEE Transactions on
  Image Processing} \bibinfo{volume}{9} (\bibinfo{year}{2000})
  \bibinfo{pages}{1484--1496}.
\bibitem[{Bustos et~al.(2009)Bustos, Flesia, Frery and
  Lucini}]{SimulationSpatiallyCorrelatedClutter2009}
\bibinfo{author}{O.H. Bustos}, \bibinfo{author}{A.G. Flesia},
  \bibinfo{author}{A.C. Frery}, \bibinfo{author}{M.M. Lucini},
  \bibinfo{title}{Simulation of spatially correlated clutter fields},
  \bibinfo{journal}{Communications in Statistics -- Simulation and Computation}
  \bibinfo{volume}{38} (\bibinfo{year}{2009}) \bibinfo{pages}{2134--2151}.
\bibitem[{Bustos et~al.(2002)Bustos, Lucini and Frery}]{Bustos2002}
\bibinfo{author}{O.H. Bustos}, \bibinfo{author}{M.M. Lucini},
  \bibinfo{author}{A.C. Frery}, \bibinfo{title}{M-estimators of roughness and
  scale for $\mathcal{G}^0_{A}$ modelled {SAR} imagery},
  \bibinfo{journal}{Eurasip Journal on Applied Signal Processing}
  \bibinfo{volume}{2002} (\bibinfo{year}{2002}) \bibinfo{pages}{105--114}.
\bibitem[{Conover(1980)}]{Conover}
\bibinfo{author}{W.J. Conover}, \bibinfo{title}{Practical nonparametric
  statistics}, \bibinfo{publisher}{John Wiley \& Sons, Inc},
  \bibinfo{year}{1980}.
\bibitem[{Cribari-Neto et~al.(2002)Cribari-Neto, Frery and
  Silva}]{Cribari-Neto2002}
\bibinfo{author}{F.~Cribari-Neto}, \bibinfo{author}{A.C. Frery},
  \bibinfo{author}{M.F. Silva}, \bibinfo{title}{Improved estimation of clutter
  properties in speckled imagery}, \bibinfo{journal}{Computational Statistics
  \& Data Analysis} \bibinfo{volume}{40} (\bibinfo{year}{2002})
  \bibinfo{pages}{801--824}.
\bibitem[{Doornik(2002)}]{Doornik98}
\bibinfo{author}{J.A. Doornik}, \bibinfo{title}{Object-Oriented Matrix
  Programming Using Ox}, \bibinfo{publisher}{Timberlake Consultants Press \&
  Oxford}, \bibinfo{address}{London}, \bibinfo{edition}{3} edition,
  \bibinfo{year}{2002}.
\bibitem[{Fesharaki and Hellestrand(1994)}]{Fesharaki1994}
\bibinfo{author}{M.N. Fesharaki}, \bibinfo{author}{G.R. Hellestrand},
  \bibinfo{title}{A new edge detection algorithm based on a statistical
  approach}, in: \bibinfo{booktitle}{International Symposium on Speech, Image
  Processing and Neural Networks (ISSIPNN)}, volume~\bibinfo{volume}{1}, pp.
  \bibinfo{pages}{21--24}.
\bibitem[{Freitas et~al.(2005)Freitas, Frery and
  Correia}]{FreitasFreryCorreia:Environmetrics:03}
\bibinfo{author}{C.C. Freitas}, \bibinfo{author}{A.C. Frery},
  \bibinfo{author}{A.H. Correia}, \bibinfo{title}{The polarimetric {G}
  distribution for {SAR} data analysis}, \bibinfo{journal}{Environmetrics}
  \bibinfo{volume}{16} (\bibinfo{year}{2005}) \bibinfo{pages}{13--31}.
\bibitem[{Frery et~al.(2004)Frery, Cribari-Neto and de~Souza}]{Frery2004}
\bibinfo{author}{A.C. Frery}, \bibinfo{author}{F.~Cribari-Neto},
  \bibinfo{author}{M.O. de~Souza}, \bibinfo{title}{Analysis of minute features
  in speckled imagery with maximum likelihood estimation},
  \bibinfo{journal}{Eurasip Journal on Applied Signal Processing}
  \bibinfo{volume}{2004} (\bibinfo{year}{2004}) \bibinfo{pages}{2476--2491}.
\bibitem[{Frery et~al.(1997)Frery, M{\"u}ller, Freitas and
  Siqueira}]{Frery1997}
\bibinfo{author}{A.C. Frery}, \bibinfo{author}{H.J. M{\"u}ller},
  \bibinfo{author}{C.C. Freitas}, \bibinfo{author}{S.J. Siqueira},
  \bibinfo{title}{A model for extremely heterogeneous clutter},
  \bibinfo{journal}{IEEE Transactions on Geoscience and Remote Sensing}
  \bibinfo{volume}{35} (\bibinfo{year}{1997}) \bibinfo{pages}{648--659}.
\bibitem[{Gambini et~al.(2008)Gambini, Mejail, Jacobo-Berlles and
  Frery}]{Gambini:StatisticsComputing}
\bibinfo{author}{J.~Gambini}, \bibinfo{author}{M.~Mejail},
  \bibinfo{author}{J.~Jacobo-Berlles}, \bibinfo{author}{A.C. Frery},
  \bibinfo{title}{Accuracy of edge detection methods with local information in
  speckled imagery}, \bibinfo{journal}{Statistics and Computing}
  \bibinfo{volume}{18} (\bibinfo{year}{2008}) \bibinfo{pages}{15--26}.
\bibitem[{Gambini et~al.(2006)Gambini, Mejail, Jacobo-Berlles and
  Frery}]{Gambini2006}
\bibinfo{author}{J.~Gambini}, \bibinfo{author}{M.E. Mejail},
  \bibinfo{author}{J.~Jacobo-Berlles}, \bibinfo{author}{A.C. Frery},
  \bibinfo{title}{Feature extraction in speckled imagery using dynamic
  {B}-spline deformable contours under the $\mathcal{G}^0$ model},
  \bibinfo{journal}{International Journal of Remote Sensing}
  \bibinfo{volume}{27} (\bibinfo{year}{2006}) \bibinfo{pages}{5037--5059}.
\bibitem[{Hoon~Lim and Ju~Jang(2002)}]{HoonLim}
\bibinfo{author}{D.~Hoon~Lim}, \bibinfo{author}{S.~Ju~Jang},
  \bibinfo{title}{Comparison of two-sample tests for edge detection in noisy
  images}, \bibinfo{journal}{Journal of the Royal Statistical Society: Series D
  (The Statistician)} \bibinfo{volume}{Volume 51 Issue 1}
  (\bibinfo{year}{2002}) \bibinfo{pages}{21--30}.
\bibitem[{Lim(2006)}]{Lim2006}
\bibinfo{author}{D.H. Lim}, \bibinfo{title}{Robust edge detection in noisy
  images}, \bibinfo{journal}{Computational Statistics \& Data Analysis}
  \bibinfo{volume}{50} (\bibinfo{year}{2006}) \bibinfo{pages}{803--812}.
\bibitem[{Mejail et~al.(2001)Mejail, Frery, Jacobo-Berlles and
  Bustos}]{Mejail2001}
\bibinfo{author}{M.E. Mejail}, \bibinfo{author}{A.C. Frery},
  \bibinfo{author}{J.~Jacobo-Berlles}, \bibinfo{author}{O.~Bustos},
  \bibinfo{title}{Approximation of distributions for {SAR} images: Proposal,
  evaluation and practical consequences}, \bibinfo{journal}{Latin American
  Applied Research} \bibinfo{volume}{31} (\bibinfo{year}{2001})
  \bibinfo{pages}{83--92}.
\bibitem[{Mejail et~al.(2003)Mejail, Jacobo-Berlles, Frery and
  Bustos}]{Mejail2003}
\bibinfo{author}{M.E. Mejail}, \bibinfo{author}{J.C. Jacobo-Berlles},
  \bibinfo{author}{A.C. Frery}, \bibinfo{author}{O.H. Bustos},
  \bibinfo{title}{Classification of {SAR} images using a general and tractable
  multiplicative model}, \bibinfo{journal}{International Journal of Remote
  Sensing} \bibinfo{volume}{24} (\bibinfo{year}{2003})
  \bibinfo{pages}{3565--3582}.
\bibitem[{Moschetti et~al.(2006)Moschetti, Palacio, Picco, Bustos and
  Frery}]{Moschetti2006}
\bibinfo{author}{E.~Moschetti}, \bibinfo{author}{M.G. Palacio},
  \bibinfo{author}{M.~Picco}, \bibinfo{author}{O.H. Bustos},
  \bibinfo{author}{A.C. Frery}, \bibinfo{title}{On the use of {Lee}'s protocol
  for speckle-reducing techniques}, \bibinfo{journal}{Latin American Applied
  Research} \bibinfo{volume}{36} (\bibinfo{year}{2006})
  \bibinfo{pages}{115--121}.
\bibitem[{Nascimento et~al.(2010)Nascimento, Cintra and
  Frery}]{NascimentoCintraFrertIEEETGARS}
\bibinfo{author}{A.D.C. Nascimento}, \bibinfo{author}{R.J. Cintra},
  \bibinfo{author}{A.C. Frery}, \bibinfo{title}{Hypothesis testing in speckled
  data with stochastic distances}, \bibinfo{journal}{IEEE Transactions on
  Geoscience and Remote Sensing} \bibinfo{volume}{48} (\bibinfo{year}{2010})
  \bibinfo{pages}{373--385}.
\bibitem[{Oliver and Quegan(1998)}]{Oliver1998}
\bibinfo{author}{C.~Oliver}, \bibinfo{author}{S.~Quegan},
  \bibinfo{title}{Understanding Synthetic Aperture Radar Images},
  \bibinfo{publisher}{Artech House, Boston.}, \bibinfo{year}{1998}.
\bibitem[{Pianto and Cribari-Neto(2011)}]{DealingMonotoneLikelihood}
\bibinfo{author}{D.M. Pianto}, \bibinfo{author}{F.~Cribari-Neto},
  \bibinfo{title}{Dealing with monotone likelihood in a model for speckled
  data}, \bibinfo{journal}{Computational Statistics and Data Analysis}
  \bibinfo{volume}{55} (\bibinfo{year}{2011}) \bibinfo{pages}{1394--1409}.
\bibitem[{Richards(2009)}]{RemoteSensingImagingSar}
\bibinfo{author}{J.A. Richards}, \bibinfo{title}{Remote Sensing with Imaging
  Radar}, \bibinfo{publisher}{Springer}, \bibinfo{year}{2009}.
\bibitem[{Su et~al.(2011)Su, He, Feng, Deng and
  Sun}]{SupervisedClassificationConditionalRandomFieldsMultiscaleRegionConnectionCalculusSAR}
\bibinfo{author}{X.~Su}, \bibinfo{author}{C.~He}, \bibinfo{author}{Q.~Feng},
  \bibinfo{author}{X.~Deng}, \bibinfo{author}{H.~Sun}, \bibinfo{title}{A
  supervised classification method based on conditional random fields with
  multiscale region connection calculus model for sar image},
  \bibinfo{journal}{IEEE Geoscience and Remote Sensing Letters}
  \bibinfo{volume}{8} (\bibinfo{year}{2011}) \bibinfo{pages}{497--501}.
\bibitem[{Vasconcellos et~al.(2005)Vasconcellos, Frery and
  Silva}]{Vasconcellos2005}
\bibinfo{author}{K.L.P. Vasconcellos}, \bibinfo{author}{A.C. Frery},
  \bibinfo{author}{L.B. Silva}, \bibinfo{title}{Improving estimation in
  speckled imagery}, \bibinfo{journal}{Computational Statistics}
  \bibinfo{volume}{20} (\bibinfo{year}{2005}) \bibinfo{pages}{503--519}.
\bibitem[{Venables and Ripley(2002)}]{VenablesRipley:S:02}
\bibinfo{author}{W.N. Venables}, \bibinfo{author}{B.D. Ripley},
  \bibinfo{title}{Modern Applied Statistics with {S}}, Statistics and
  Computing, \bibinfo{publisher}{Springer}, \bibinfo{address}{New York},
  \bibinfo{edition}{4} edition, \bibinfo{year}{2002}.

\end{thebibliography}

\label{lastpage}

\end{document}